\documentclass[12pt,letterpaper]{article}
\usepackage{graphicx,amsfonts,amsmath,amssymb,hyperref,float,setspace}
\usepackage[font=small]{caption,subcaption}
\numberwithin{equation}{section}

  \renewenvironment{thebibliography}[1]{
   \begin{oldthebibliography}{#1}
    \setlength{\itemsep}{-0.3mm}       }
     {\end{oldthebibliography}}
\begin{document}
\begin{titlepage}

\hbox to \hsize{\hspace*{0 cm}\hbox{\tt }\hss
    \hbox{\small{\tt }}}

\vskip1cm
\centerline{\Large\bf Perturbatively Charged  Holographic Disorder}

\vspace{1 cm}
 \centerline{\large Daniel K. O'Keeffe\footnote{dokeeffe@physics.utoronto.ca}\, and Amanda W. Peet\footnote{awpeet@physics.utoronto.ca}}

\vspace{0.3cm}
\begin{center}
{\it Department of Physics, \\{} University of Toronto, \\{} Toronto, Ontario, \\{} Canada M5S 1A7.}
\end{center}
 
\vspace{1 cm}

\begin{abstract}

Within the framework of holography applied to condensed matter physics, we study a model of perturbatively charged disorder in $D=4$ dimensions.  Starting from initially uncharged $AdS_{4}$, a randomly fluctuating boundary chemical potential is introduced by turning on a bulk gauge field parameterized by a disorder strength $\overline{V}$ and a characteristic scale $k_{0}$.  Accounting for gravitational backreaction, we construct an asymptotically AdS solution perturbatively in the disorder strength. The disorder averaged geometry displays unphysical divergences in the deep interior. We explain how to remove these divergences and arrive at a well behaved solution.  The disorder averaged DC conductivity is calculated and is found to contain a correction to the AdS result.  The correction appears at second order in the disorder strength and scales inversely with $k_{0}$. We discuss the extension to a system with a finite initial charge density.  The disorder averaged DC conductivity may be calculated by adopting a technique developed for holographic lattices.  

\end{abstract}

\end{titlepage}

\begin{spacing}{0.9}
\tableofcontents
\end{spacing}

\section{Introduction}\label{sec:intro}

The AdS/CFT correspondence has proven to be a remarkably powerful tool for probing the detailed structure of strongly coupled quantum field theories.  Its broad list of successes includes applications to modelling the quark-gluon plasma, condensed matter phenomena such as superconductivity, and even fluid dynamics.  Comprehensive reviews of these subjects include, \cite{Casalderrey-Solana2011}, \cite{Cai2015} and \cite{Hubeny2012}, respectively.  

An underlying theme to this progress is the reduction of symmetry in holographic models.  Systems which possess too many symmetries display behaviours that are not desirable in condensed matter models, a prime example being the infinite DC conductivity dual to the Reissner-Nordstr\"{o}m-AdS (RN-AdS) geometry.  This is not a surprising feature. The underlying RN-AdS geometry possesses translational invariance along the boundary directions. This means that the charge carriers in the dual theory have no means by which to dissipate momentum, resulting in an infinite DC conductivity. Realistic condensed matter systems do not display this behaviour, so if a holographic model is going to be useful for studying these kinds of problems, we need a way to break translational invariance.

Several avenues of investigation have been carried out, including explicit holographic lattices: \cite{Kachru2010}, \cite{Kachru2011}, \cite{Horowitz2012}, \cite{Horowitz2012a}, \cite{Horowitz2013a}, \cite{Blake2013}, \cite{Hartnoll2014a}, \cite{Donos2015}, Q-lattices: \cite{Donos2014}, \cite{Donos2014b}, \cite{Ling2015} and breaking diffeomorphism invariance: \cite{Andrade2014b}, \cite{Gouteraux2014a}, \cite{Mefford2014a}, \cite{Taylor2014}, \cite{Withers2014}, \cite{Kim2014}, \cite{Davison2015}, \cite{Andrade2014c}.  Other examples of holographic symmetry breaking include the breaking of rotational symmetry \cite{Iizuka2012}, \cite{Iizuka2013}, \cite{Donos2012b}, \cite{Donos2014c}, \cite{Erdmenger2015}, relativistic symmetry \cite{Son2008}, \cite{Balasubramanian2008}, \cite{Kachru2008} and hyperscaling violation \cite{Gouteraux2011}, \cite{Ogawa2012}, \cite{Huijse2012}, \cite{Dong2012}.  Our focus will be on disorder. Within the context of bottom-up modelling, we will report here on a perturbative construction of holographic disorder.  That is, we will seek a particular spacetime solution that is sourced by a disordered, randomly fluctuating field in the dual theory and use it to study the dual DC conductivity.  

Disorder is a common feature in real world condensed matter systems, but it is difficult to model using traditional field theory approaches: little is known at strong coupling. This is especially true in the context of localization.  It is well known that in a non-interacting system, the addition of disorder can completely suppress conductivity \cite{Anderson1958}; for a comprehensive review, see \cite{Lee1985}.  Turning on interactions complicates the situation and many-body localization may occur \cite{Basko2006}; a review may be found in \cite{Nandkishore2014}. Experimental studies of many-body localization are also in their infancy. Since holography is a strong-weak equivalence, it provides a potential way forward for studying disorder. Understanding what the gravity dual tells us about localization might even provide a definition of many body localization.  Exactly what this would look like in a gravitational theory and how the bulk fields would behave is an open and interesting question.  

Several approaches to disorder within holography have been proposed.  See \cite{Hartnoll2007}, \cite{Hartnoll2008} and \cite{Ryu2011} for early studies.   The replica trick for disordered CFTs was extended to holography in \cite{Fujita2008} and later in \cite{Shang2012}.

To fully see the effect of quenched random disorder, \cite{Adams2011} proposed that disorder should be modelled by applying random boundary conditions to a bulk field and allowing the field to backreact on the geometry.  The source of disorder is characterized by a distribution $P_{V}[W(x)]$, over random functions $W(x)$.  The subscript $V$ is in reference to another function $V(x)$, which is taken to be the boundary value of a gauge field in the bulk.  The system is assumed to be self-averaging, meaning that over length scales much larger than than a typical disorder scale, homogeneity is restored.  An interesting feature of this construction is that the entire functional $P_{V}[W(x)]$ runs as the energy scale changes.  To this end, \cite{Adams2011} and later \cite{Adams2014}, worked to construct a holographic functional renormalization scheme so that disorder averaged thermodynamic quantities could be computed.   

Another approach to disorder in gauge/gravity duality was proposed in \cite{Saremi2014}.  A particular background which is already deformed away from AdS, but still satisfies the null energy condition, is considered. The degree to which the geometry is deformed is controlled by a parameter which is meant to represent the amount of disorder in the system. The wave function of a probe scalar in this background is studied by looking at the nearest neighbour spacing distribution of the pole spectra of the two-point correlator. By increasing the amount of disorder, a transition is observed between an initial distribution and a Poisson distribution, which is analogous to what happens in a disorder driven metal-insulator transition.  

More recently, \cite{Arean2014} applied a spectral approach to modelling disorder in a holographic superconductor.  The basic idea is similar to \cite{Adams2011} where the disorder is sourced by a random space-dependent chemical potential by setting the boundary condition on a bulk $U(1)$ gauge field.  In this proposal, the chemical potential takes the form of a sum over a spectrum which depends on one of the boundary directions, $x$
\begin{equation}{\label{eqn:introchemdis}}
\mu(x) = \mu_{0} + \overline{V} \displaystyle\sum_{k = k_{0}}^{k_{\ast}} \sqrt{S_{k}} \cos(k x + \delta_{k}) \, ,
\end{equation}
where $S_{k}$ is a function of the momenta $k$ which controls the correlation function for $\mu(x)$; different choices of $S_{k}$ lead to a different values of the disorder distribution average.  $\overline{V}$ is a parameter which sets the strength of the disorder.  $k_{0}$ and $k_{\ast}$ define an IR and UV length scale cutoff for the disorder, respectively.  $\delta_{k}$ is a random phase for each value of $k$.  A spectral representation like this is known to simulate a stochastic process when $k_{\ast}$ is large \cite{Shinozuka1991}.

The dirty chemical potential (\ref{eqn:introchemdis}) is incorporated into an Einstein-Maxwell model.  The Maxwell equations are solved numerically in an electric ansatz with the boundary condition that $A_{t}$ approaches (\ref{eqn:introchemdis}) near the boundary.  The analysis is done in a fixed AdS-Schwarzschild background.  Evidence is found for an enhancement of the critical temperature of the superconductor for increasing disorder strength.  Numerical evidence that localization occurs in this model was reported in \cite{Zeng2013}.  This model has also been extended to holographic p-wave superconductors in \cite{Arean2014a} where the same behaviour of the critical temperature with disorder strength was observed.  

There is a resemblance between this spectral approach to disorder and the construction of a holographic ionic lattice in \cite{Horowitz2012} and \cite{Chesler2014} where the lattice is set up by a periodic boundary chemical potential.  The difference is that for a holographic lattice, there is only a single periodic source of a fixed wavelength. In the spectral approach to disorder, there is a sum over periodic sources of arbitrary wavelength, so the effect of disorder may resonate more strongly throughout the entire bulk geometry, having non-trivial effects deep in the interior.  

Recently, \cite{Hartnoll2014} applied a spectral approach to modelling disorder sourced by a scalar field in $2+1$ bulk dimensions.  The initial clean geometry in this case is $AdS_{3}$ and the scalar field is allowed to backreact on the geometry in the spirit of \cite{Adams2011}.  By treating the disorder strength as a perturbative handle, a second order analytic solution for the backreacted geometry is obtained.  It is observed that the disorder averaged geometry, in the deep interior, takes the form of a Lifshitz metric with a dynamical critical exponent $z$ set by the disorder strength.  Numerical solutions are constructed for strong disorder, also display this behaviour.  A similar implementation of scalar disorder was used in \cite{Lucas2014} and \cite{Lucas2015a} to study the conductivity of holographic strange metals with weak quenched disorder.  

In \cite{Hartnoll2014a}, it was shown that a wide variety of inhomogeneous IR geometries arise as solutions to Einstein-Maxwell theory with a single periodic source at the boundary.  These observations suggest that studying disordered boundary potentials in Einstein-Maxwell theory at both zero and finite net charge density may lead to novel gravitational solutions and insight into disordered condensed matter systems. 

Our goal here is to study charged disorder with backreaction. As we will see, the general problem for a system with baseline charge density is very complicated. Accordingly, we study the story perturbatively.  We build in the disorder the same way as \cite{Arean2014}, \cite{Arean2014a}, with a randomly varying boundary chemical potential, and we also  include backreaction. In the bulk, this corresponds to having a fluctuating gauge field and letting it backreact on the initially clean AdS geometry.  We then solve the Einstein equations perturbatively in the strength of the disorder and construct an analytic, asymptotically AdS, solution. Similar to the case of scalar disorder studied in \cite{Hartnoll2014}, the disorder averaged geometry contains unphysical secular terms which diverge in the deep interior.  We explain how these divergences may be tamed and ultimately find a well behaved averaged geometry.

Our primary interest is in investigating the transport properties dual to the disordered geometry.  Adapting a technique first developed in \cite{Donos2014b}, we directly calculate the disorder averaged DC conductivity. We find a correction to the conductivity dual to pure AdS starting at second order in the disorder strength which scales inversely with the smallest wavenumber $k_{0}$ in the disordered chemical potential.  We will also discuss extensions to systems with finite charge density.    

Holographic momentum dissipation and disorder has also been approached with the context of massive gravity.  See \cite{Vegh2013}, \cite{Davison2013a}, \cite{Blake2013a}, \cite{Amoretti2014}, \cite{Amoretti2015a}, \cite{Baggioli2014}, and \cite{Amoretti2015}. While this approach will not be our primary focus here, it will nevertheless be interesting to understand how results within massive gravity mesh with explicit implementations of holographic disorder. 

This paper is organized as follows.  In section \ref{sec:uncharged} we construct a bottom-up, disordered, holographic spacetime analytically.  Using the strength of the disorder as a perturbative handle, we solve the bulk equations of motion up to second order.  Using a Gaussian distribution for the disorder source, we explicitly evaluate the disorder average of the backreacted solution. In section \ref{sec:resum}, we implement a resummation procedure for removing spurious secular terms which develop in the disorder averaged geometry as a result of our perturbative expansion.  With the regulated solution in hand, we study the resultant DC conductivity in section \ref{sec:cond}.  Section \ref{sec:Finite} points out some features of incorporating disorder in a geometry with an initial charge density. We point out how techniques used to study transport properties of holographic lattices may be used to access the DC conductivity in the finite charge case.  Finally, in section \ref{sec:discussion} we summarize our findings and comment on possible directions for future work.

\section{Perturbatively charged disorder}\label{sec:uncharged}

We are interested in studying the effect of a disordered holographic lattice, with the ultimate goal of understanding transport properties like conductivity in the dual model.  To this end, we start with a gauge field the $D= d+1$ dimensional bulk and then introduce a random perturbation around the initially clean baseline solution.  The model is then Einstein-Maxwell gravity
\begin{equation}{\label{eqn:action}}
S = \frac{1}{16 \pi G_{N}} \displaystyle\int d^{d+1} x \sqrt{-g} \left ( R + \frac{d(d-1)}{L^{2}} - \frac{1}{4} F^{2} \right ) \, .
\end{equation}
\noindent Note that since we want an asymptotically AdS solution, we have set the cosmological constant to $\Lambda = -d(d-1)/2 L^{2}$, where $L$ is the AdS radius.  The equations of motion we need to solve are then the Einstein and Maxwell equations.  In what follows, it will be convenient to work with the traced Einstein equations, that is we need to solve
\begin{equation}{\label{eqn:Ein}} 
R_{\mu \nu} + \frac{d}{L^{2}} g_{\mu \nu} = \frac{1}{2} \left ( F_{\mu}^{\sigma} F_{\nu \sigma} - \frac{1}{2(d-1)} g_{\mu \nu} F^{2} \right ) \, ,
\end{equation}
\begin{equation}{\label{eqn:Max}}
\nabla_{\mu} F^{\mu \nu} = 0 \, .
\end{equation}
We consider a system where the initial charge density is zero.  In this case, the baseline solution is just $AdS_{d+1}$ with $A_{t} = 0$.  That is, the initially clean (no disorder) background is
\begin{equation}{\label{eqn:AdS}}
ds^{2} = \frac{L^{2}}{r^{2}} \left ( -dt^{2} + dr^{2} + dx_{i}^{2} \right ) \, ,
\end{equation}
\noindent where the boundary is at $r=0$ in these coordinates.  To introduce disorder, we will perturb about this background by turning on the gauge field.  To get a tractable system, we will introduce disorder along only one of the boundary direction $x$.   According to the holographic dictionary, near the boundary, the time component of the gauge field behaves like 
\begin{equation}{\label{eqn:GaugeFG}}
A_{t}(x,r) = \mu(x) + \rho(x) r^{d-2} + \cdots \, ,
\end{equation}
where $\mu(x)$ is the chemical potential and $\rho(x)$ is related to the charge density of the dual theory.  We source the disorder by taking the chemical potential to be a sum of periodic functions with random phases using a spectral representation
\begin{equation}{\label{eqn:chemicalpotential}}
\mu(x) = \overline{V} \displaystyle\sum_{n=1}^{N-1} A_{n} \cos(k_{n} x + \theta_{n}) \, .
\end{equation}
Here, $\overline{V}$ is a constant which controls the strength of the disorder which we will use as a perturbative handle. Similar profiles for the gauge field have been studied in the context of disordered holographic superconductors in \cite{Arean2014}, \cite{Zeng2013}, \cite{Arean2014a}.  A spectral representation for scalar disorder was used in \cite{Hartnoll2014}.  The wavenumbers $k_{n} = n \Delta k$ are evenly spaced with $\Delta k = k_{0}/N$.  $1/k_{0}$ is thought of as a short distance cutoff for the disorder and is held fixed so that in the limit that $N \rightarrow \infty$, $\Delta k \rightarrow 0$.  The $\theta_{n}$ are random angles that are assumed to be uniformly distributed in $[0,2\pi]$.  The amplitudes are 
\begin{equation}{\label{eqn:Amp}}
A_{n} = 2 \sqrt{S(k_{n}) \Delta k} \, ,
\end{equation}
where $S(k_{n})$ controls the correlation functions.  Since the $\theta_{n}$ are uniformly distributed, averaging over them (the disorder average) is done simply by taking
\begin{equation}{\label{eqn:average}}
\langle f \rangle _{D} = \lim_{N \rightarrow + \infty} \displaystyle\int \displaystyle\prod_{i=1}^{N-1} \frac{d \theta_{i}}{2 \pi} f \, .
\end{equation}
If the function $S(k_{n})$ is taken to be 1, this leads to $\mu(x)$ describing Gaussian noise.  In other words, by explicit calculation, we have that
\begin{equation}{\label{eqn:Gaussian}}
\langle \mu(x) \rangle_{D} = 0 \, , \quad \langle \mu(x_{1}) \mu(x_{2}) \rangle_{D} = \overline{V}^{\, 2} \delta(x_{1} - x_{2}) \, .
\end{equation}
The gauge field $A_{t}$ sources the disorder in the system, which means that it contributes at order $\overline{V}$.  This in turn implies that the right hand side of the traced Einstein equations (\ref{eqn:Ein}) start contributing at second order in the disorder strength.  Only Maxwell's equations receive any corrections at first order in the disorder strength, so we can solve for $A_{t}$ in the clean $AdS$ background.  The solution which obeys the boundary condition (\ref{eqn:chemicalpotential}) at $r=0$ is
\begin{equation}{\label{eqn:firstordermax}}
A_{t}(r,x) = \overline{V} \displaystyle\sum_{n=1}^{N-1} \frac{A_{n} k_{n}^{(d-2)/2}}{2^{(d-4)/2} \Gamma(\frac{d-2}{2})} r^{(d-2)/2} K_{(d-2)/2}(k_{n}r) \cos(k_{n} x + \theta_{n}) \, ,
\end{equation}
\noindent where $K_{(d-2)/2}(k_{n}r)$ is the modified Bessel function of the second kind.  Notice, by turning on disorder, we went from a solution that had zero charge density to one that has a finite charge density.  We can compute this in the dual theory: it is $\langle J^{t} \rangle$, i.e. the expectation value of the current density.  The result is
\begin{equation}{\label{eqn:DisorderCurrent}}
\langle J^{t} \rangle = -\frac{(d-2) \overline{V} L^{d-3}}{8 \pi G_{N}} \displaystyle\sum_{n=1}^{N-1} \frac{A_{n} k_{n}^{d-2}}{2^{d-1}} \frac{\Gamma((2-d)/2)}{\Gamma((d-2)/2)} \cos(k_{n} x + \theta_{n}) \, .
\end{equation}
\noindent Note, this formula is valid when $(2-d)/2$ is not a negative integer (which will be the case in what follows).  Now, we can plug the solution for $A_{t}$ back into the traced Einstein equations and work out the metric corrections at second order in $\overline{V}$.  This is generally difficult to do in arbitrary dimensions.  We will focus on the case with $D = 4$, in this way the dual to the initially clean geometry is a $2+1$ dimensional CFT.  We need an ansatz for the backreacted metric.  Since we are adding a perturbation in the $x$ direction only, the metric coefficients could generally depend on both the radial direction $r$ and the boundary direction $x$.  Furthermore, we insist that the geometry is asymptotically AdS and we will work within Fefferman-Graham gauge.  At the end of the day, this means that the general form of the backreacted metric is
\begin{equation}{\label{eqn:Backreact}}
ds^{2} = \frac{L^{2}}{r^{2}} \left (-\alpha(x,r) dt^{2} + dr^{2} + \eta(x,r) dx^{2} + \delta(x,r) dy^{2} \right ) \, ,
\end{equation}
\noindent where the functions $\alpha(x,r)$, $\eta(x,r)$ and $\delta(x,r)$ need to be solved for.  We solve for them in a perturbative expansion
\begin{align*}
&\alpha(x,r)=1+\overline{V}^{\, 2} \alpha_{2}(x,r) + \cdots \\
&\eta(x,r)= 1+\overline{V}^{\, 2} \eta_{2}(x,r) + \cdots \\
&\delta(x,r)=1+\overline{V}^{\, 2} \delta_{2}(x,r) + \cdots \, ,
\end{align*}
\noindent so to second order we solve for $\alpha_{2}(x,r)$, $\eta_{2}(x,r)$ and $\delta_{2}(x,r)$.  The traced Einstein equations (\ref{eqn:Ein}) expanded to second order give the following set of equations for the metric coefficient 
\begin{equation}\label{eqn:eom1tt}
-2r \partial_{r} \delta_{2} + 2 r^{2} \partial_{x}^{2} \alpha_{2} - 2 r \partial_{r} \eta_{2} + 2 r^{2} \partial_{r}^{2} \alpha_{2} - 6 r \partial_{r} \alpha_{2} = \frac{r^{4}}{L^{2}} \left [ (\partial_{r} H)^{2} + (\partial_{x} H)^{2} \right ] \, ,
\end{equation}
\begin{equation}\label{eqn:eom1rr}
-2 r^{2} \partial_{r}^{2} \delta_{2} + 2 r \partial_{r} \eta_{2} + 2 r \partial_{r} \alpha_{2} - 2 r^{2} \partial_{r}^{2} \alpha_{2} - 2 r^{2} \partial^{2}_{r} \eta_{2} + 2 r \partial_{r} \delta_{2} = - \frac{r^{4}}{L^{2}} \left [ (\partial_{r} H)^{2} - (\partial_{x} H)^{2} \right ] \, ,
\end{equation}
\begin{equation}\label{eqn:eom1rx}
\partial_{r} \partial_{x} (\alpha_{2} + \delta_{2} ) = \frac{r^{2}}{L^{2}} (\partial_{r} H) (\partial_{x} H) \, ,
\end{equation}
\begin{equation}\label{eqn:eom1xx}
6 r \partial_{2} \eta_{2} + 2 r \partial_{r} \alpha_{2} + 2 r \partial_{r} \delta_{2} - 2 r^{2} \partial_{x}^{2} \alpha_{2} - 2 r^{2} \partial_{x}^{2} \delta_{2} - 2 r^{2} \partial_{r}^{2} \eta_{2} = \frac{r^{4}}{L^{2}} \left [ (\partial_{r} H)^{2} - (\partial_{x} H)^{2} \right ] \, ,
\end{equation}
\begin{equation}\label{eqn:eom1yy}
2 r \partial_{r} \eta_{2} + 2 r \partial_{r} \alpha_{2} + 6 r \partial_{r} \delta_{2} - 2 r^{2} \partial_{r}^{2} \delta_{2} - 2 r^{2} \partial_{x}^{2} \delta_{2} = \frac{r^{4}}{L^{2}} \left [ (\partial_{r} H)^{2} + (\partial_{x} H)^{2} \right ] \, ,
\end{equation}
\noindent where $H = H(x,r)$ such that $A_{t} = \overline{V} H(x,r)$ is the solution to the Maxwell equations at first order (\ref{eqn:firstordermax}).

\subsection{Second order solution}\label{sec:sol2}
By using combinations of equations (\ref{eqn:eom1tt})-(\ref{eqn:eom1yy}), we can solve for $\alpha_{2}(x,r)$, $\eta_{2}(x,r)$ and $\delta_{2}(x,r)$.  The solutions are
\begin{align*}
&\alpha_{2}(x,r) =  \frac{1}{2} H_{1}(x)+\frac{1}{2} c_{1} + \frac{1}{2} c_{2} r^{3} - \frac{1}{16 L^{2}} \displaystyle\sum_{n=1}^{N-1} \frac{A_{n}^{2}}{k_{n}^{2}} + \frac{1}{32 L^{2}} \displaystyle\sum_{n=1}^{N-1} \frac{A_{n}^{2}}{k_{n}^{2}} (2 k_{n}^{2} r^{2} + 2 k_{n} r +1) e^{-2 k_{n} r} \\
&+\frac{1}{16 L^{2}} \displaystyle\sum_{n=1}^{N-1} \frac{A_{n}^{2}}{k_{n}^{2}} (2 k_{n}^{2} r^{2} + 2 k_{n} r +1) \exp(-2 k_{n} r) \cos^{2}(k_{n} x + \theta_{n}) \\
&+ \frac{1}{8 L^{2}} \displaystyle\sum_{n=1}^{N-1} \displaystyle\sum_{m=1}^{n-1} \frac{A_{n} A_{m}}{k_{n} k_{m}} \left [1 + (k_{n} + k_{m})r + 2 k_{n} k_{m} r^{2} \right ] e^{-(k_{n} + k_{m}) r} \cos((k_{n} - k_{m}) x + \theta_{n} - \theta_{m}) \\
&-\frac{1}{8 L^{2}} \displaystyle\sum_{n=1}^{N-1} \displaystyle\sum_{m=1}^{n-1} \frac{A_{n} A_{m}}{k_{n} k_{m}} \left [ 1 + (k_{n} - k_{m})r \right ] e^{-(k_{n} - k_{m})r} \cos((k_{n}-k_{m})x + \theta_{n} - \theta_{m}) \\
&+\frac{1}{2 L^{2}} \displaystyle\sum_{n=1}^{N-1} \displaystyle\sum_{m=1}^{n-1} \frac{A_{n} A_{m} k_{n} k_{m}}{(k_{n} + k_{m})^{4}} \left [ (k_{n}+k_{m})^{2} r^{2} + 2 (k_{n} + k_{m}) r + 2 \right ] e^{-(k_{n} + k_{m})r} \\
&\hskip 12em \times \cos((k_{n} + k_{m}) x + \theta_{n} + \theta_{m}) \, ,
\end{align*}
\begin{align*}
&\delta_{2}(x,r)  =  \frac{1}{2} H_{1}(x)+\frac{1}{2} c_{1}  + \frac{1}{2} c_{2} r^{3} + \frac{1}{16 L^{2}} \displaystyle\sum_{n=1}^{N-1} \frac{A_{n}^{2}}{k_{n}^{2}} - \frac{3}{32 L^{2}} \displaystyle\sum_{n=1}^{N-1} \frac{A_{n}^{2}}{k_{n}^{2}} (2 k_{n}^{2} r^{2} + 2 k_{n} r + 1) e^{-2 k_{n} r} \\
&+\frac{1}{16 L^{2}} \displaystyle\sum_{n=1}^{N-1} \frac{A_{n}^{2}}{k_{n}^{2}} (2 k_{n}^{2} r^{2} + 2 k_{n} r +1) \exp(-2 k_{n} r) \cos^{2}(k_{n} x + \theta_{n}) \\
&-\frac{1}{8 L^{2}} \displaystyle\sum_{n=1}^{N-1} \displaystyle\sum_{m=1}^{n-1} \frac{A_{n} A_{m}}{k_{n} k_{m}} \left [ 1 + (k_{n} + k_{m})r + 2 k_{n} k_{m} r^{2} \right ] e^{-(k_{n} + k_{m})r} \cos((k_{n} - k_{m})x + \theta_{n} - \theta_{m}) \\
&+\frac{1}{8 L^{2}} \displaystyle\sum_{n=1}^{N-1} \displaystyle\sum_{m=1}^{n-1} \frac{A_{n} A_{m}}{k_{n} k_{m}} \left [ 1 -(k_{n} - k_{m})r \right ] e^{-(k_{n} - k_{m})r} \cos((k_{n} - k_{m})x + \theta_{n} - \theta_{m}) \\
&+\frac{1}{2 L^{2}} \displaystyle\sum_{n=1}^{N-1} \displaystyle\sum_{m=1}^{n-1} \frac{A_{n} A_{m} k_{n} k_{m}}{(k_{n} + k_{m})^{4}} \left [ (k_{n} + k_{m})^{2} r^{2} + 2 (k_{n} + k_{m}) r + 2 \right ] e^{-(k_{n} + k_{m})r} \\
& \hskip 12em \times \cos((k_{n} + k_{m})x + \theta_{n} - \theta_{m}) 
\end{align*}
\begin{align*}
&\eta_{2}(x,r) = \frac{1}{4} r^{2} \partial^{2}_{x} H_{1}(x) - c_{2} r^{3} + H_{3}(x)  - \frac{1}{32 L^{2}} \displaystyle\sum_{n=1}^{N-1} \frac{A_{n}^{2}}{k_{n}^{2}} ( 2 k_{n} r + 1) \exp(-2 k_{n} r) \\
&+ \frac{1}{16 L^{2}} \displaystyle\sum_{n=1}^{N-1} \frac{A_{n}^{2}}{k_{n}^{2}} ( 2 k_{n} r + 1) \exp(-2 k_{n} r) \cos^{2}(k_{n} x + \theta_{n}) \\
&+\frac{1}{L^{2}} \displaystyle\sum_{n=1}^{N-1} \displaystyle\sum_{m=1}^{n-1} \frac{A_{n} A_{m} k_{n} k_{m}}{(k_{n} + k_{m})^{4}} \left [ (k_{n} + k_{m})r + 1 \right ] \exp(-(k_{n} + k_{m})r) \cos((k_{n} + k_{m})x + \theta_{n} + \theta_{m}) \, ,
\end{align*}
\noindent where $c_{1}$ and $c_{2}$ are constants and $H_{1}(x)$ and $H_{3}(x)$ are arbitrary functions of $x$.    The solution for the metric coefficient in the $x$ direction, $\eta_{2}(x,r)$, is the simplest, as the equations of motion (\ref{eqn:eom1tt})-(\ref{eqn:eom1yy}) can be arranged to give a first order differential equation for $\eta_{2}(x,r)$.  

We want a solution that is asymptotically $AdS$ as $r \rightarrow 0$ and is regular in the interior.  The former condition is tantamount to insisting that $\alpha_{2}(x,0) = \eta_{2}(x,0) = \delta_{2}(x,0)$ so that the metric takes on the correct asymptotically $AdS$ form in Fefferman-Graham gauge.  Imposing these conditions fixes $c_{2} = 0$ and $H_{1}(x) = \rm{const}$ and $H_{3}(x) = \rm{const}$.  The solutions for $\alpha_{2}(x,r)$, $\eta_{2}(x,r)$ and $\delta_{2}(x,r)$ which are asymptotically $AdS$ are regular in the interior are
\begin{align*}
&\alpha_{2}(x,r) =  \frac{1}{2} c - \frac{1}{16 L^{2}} \displaystyle\sum_{n=1}^{N-1} \frac{A_{n}^{2}}{k_{n}^{2}} + \frac{1}{32 L^{2}} \displaystyle\sum_{n=1}^{N-1} \frac{A_{n}^{2}}{k_{n}^{2}} (2 k_{n}^{2} r^{2} + 2 k_{n} r +1) \exp(-2 k_{n} r) \\
&+\frac{1}{16 L^{2}} \displaystyle\sum_{n=1}^{N-1} \frac{A_{n}^{2}}{k_{n}^{2}} (2 k_{n}^{2} r^{2} + 2 k_{n} r +1) \exp(-2 k_{n} r) \cos^{2}(k_{n} x + \theta_{n}) \\
&+ \frac{1}{8 L^{2}} \displaystyle\sum_{n=1}^{N-1} \displaystyle\sum_{m=1}^{n-1} \frac{A_{n} A_{m}}{k_{n} k_{m}} \left [1 + (k_{n} + k_{m})r + 2 k_{n} k_{m} r^{2} \right ] e^{-(k_{n} + k_{m}) r} \cos((k_{n} - k_{m}) x + \theta_{n} - \theta_{m}) \\
&-\frac{1}{8 L^{2}} \displaystyle\sum_{n=1}^{N-1} \displaystyle\sum_{m=1}^{n-1} \frac{A_{n} A_{m}}{k_{n} k_{m}} \left [ 1 + (k_{n} - k_{m})r \right ] e^{-(k_{n} - k_{m})r} \cos((k_{n}-k_{m})x + \theta_{n} - \theta_{m}) \\
&+\frac{1}{2 L^{2}} \displaystyle\sum_{n=1}^{N-1} \displaystyle\sum_{m=1}^{n-1} \frac{A_{n} A_{m} k_{n} k_{m}}{(k_{n} + k_{m})^{4}} \left [ (k_{n}+k_{m})^{2} r^{2} + 2 (k_{n} + k_{m}) r + 2 \right ] e^{-(k_{n} + k_{m})r} \\
&\hskip 12em \times \cos((k_{n} + k_{m}) x + \theta_{n} + \theta_{m}) \, ,
\end{align*}
\begin{align*}
&\delta_{2}(x,r)  =  \frac{1}{2} c + \frac{1}{16 L^{2}} \displaystyle\sum_{n=1}^{N-1} \frac{A_{n}^{2}}{k_{n}^{2}} - \frac{3}{32 L^{2}} \displaystyle\sum_{n=1}^{N-1} \frac{A_{n}^{2}}{k_{n}^{2}} (2 k_{n}^{2} r^{2} + 2 k_{n} r + 1) \exp(-2 k_{n} r) \\
&+\frac{1}{16 L^{2}} \displaystyle\sum_{n=1}^{N-1} \frac{A_{n}^{2}}{k_{n}^{2}} (2 k_{n}^{2} r^{2} + 2 k_{n} r +1) \exp(-2 k_{n} r) \cos^{2}(k_{n} x + \theta_{n}) \\
&-\frac{1}{8 L^{2}} \displaystyle\sum_{n=1}^{N-1} \displaystyle\sum_{m=1}^{n-1} \frac{A_{n} A_{m}}{k_{n} k_{m}} \left [ 1 + (k_{n} + k_{m})r + 2 k_{n} k_{m} r^{2} \right ] e^{-(k_{n} + k_{m})r} \cos((k_{n} - k_{m})x + \theta_{n} - \theta_{m}) \\
&+\frac{1}{8 L^{2}} \displaystyle\sum_{n=1}^{N-1} \displaystyle\sum_{m=1}^{n-1} \frac{A_{n} A_{m}}{k_{n} k_{m}} \left [ 1 -(k_{n} - k_{m})r \right ] e^{-(k_{n} - k_{m})r} \cos((k_{n} - k_{m})x + \theta_{n} - \theta_{m}) \\
&+\frac{1}{2 L^{2}} \displaystyle\sum_{n=1}^{N-1} \displaystyle\sum_{m=1}^{n-1} \frac{A_{n} A_{m} k_{n} k_{m}}{(k_{n} + k_{m})^{4}} \left [ (k_{n} + k_{m})^{2} r^{2} + 2 (k_{n} + k_{m}) r + 2 \right ] e^{-(k_{n} + k_{m})r} \\
& \hskip 12em \times \cos((k_{n} + k_{m})x + \theta_{n} - \theta_{m}) \, ,
\end{align*}
\begin{align*}
&\eta_{2}(x,r) = c  - \frac{1}{32 L^{2}} \displaystyle\sum_{n=1}^{N-1} \frac{A_{n}^{2}}{k_{n}^{2}} ( 2 k_{n} r + 1) \exp(-2 k_{n} r) \\
&+ \frac{1}{16 L^{2}} \displaystyle\sum_{n=1}^{N-1} \frac{A_{n}^{2}}{k_{n}^{2}} ( 2 k_{n} r + 1) \exp(-2 k_{n} r) \cos^{2}(k_{n} x + \theta_{n}) \\
&+\frac{1}{L^{2}} \displaystyle\sum_{n=1}^{N-1} \displaystyle\sum_{m=1}^{n-1} \frac{A_{n} A_{m} k_{n} k_{m}}{(k_{n} + k_{m})^{4}} \left [ (k_{n} + k_{m})r + 1 \right ] \exp(-(k_{n} + k_{m})r) \cos((k_{n} + k_{m})x + \theta_{n} + \theta_{m}) \, ,
\end{align*}
\noindent where $c$ is a constant.  The fact an asymptotically $AdS$ solution exists was to be expected as in $d+1=4$ dimensions, the gauge field which encodes the disorder sources a relevant operator in the dual theory.  Geometrically, the gauge field falls off sufficiently fast as $r \rightarrow 0$ so that it does not disrupt the asymptotic form of the metric.    

\subsection{Disorder average}\label{sec:avg}

The solutions we have found for $\alpha_{2}(x,r)$, $\eta_{2}(x,r)$ and $\delta_{2}(x,r)$ simplify considerably under disorder averaging. Using (\ref{eqn:average}), we average over the random angles $\theta_{n}$.  The resulting sums may be expressed in terms of special functions
\begin{align}{\label{eqn:Nalpha2}}
&\langle \alpha_{2} \rangle_{D} = \lim_{N \rightarrow + \infty} \left \{ \frac{1}{2}c - \frac{1}{4 k_{0} L^{2}} \left [ N H_{(N-1)}^{(2)} + 2 k_{0} r \ln(1-\exp(-2k_{0} r /N)) \right . \right . \\
&\left . \left . + 2 k_{0} r \exp(-2 k_{0} r) \Phi( \exp(-2k_{0} r/N) ,1,N)  - N Li_{2} \left[ \exp(-2k_{0}r/N) \right ] \right. \Big ] \right . \notag \\
&\left . \left . + \frac{2 k_{0}^{2} r^{2}}{N} \frac{\exp(-2 k_{0} r(N-1)/N) - 1}{\exp(2k_{0} r/N) -1} + N \exp(-2 k_{0} r) \Phi( \exp(-2k_{0}r/N),2,N) \right . \right . 
\bigg \} \, , \notag
\end{align}
\begin{align}{\label{eqn:Ndelta2}}
&\langle \delta_{2} \rangle_{D} = \lim_{N \rightarrow + \infty} \left \{ \frac{1}{2}c + \frac{1}{4 k_{0} L^{2}} N H_{(N-1)}^{(2)} - \frac{1}{4 k_{0} L^{2}} \left. \Big [ -2 k_{0} r \ln(1-\exp(-2k_{0} r /N)) \right . \right . \\
&\left . \left . - 2 k_{0} r \exp(-2 k_{0} r) \Phi( \exp(-2k_{0} r/N) ,1,N) + N Li_{2} \left[ \exp(-2k_{0}r/N) \right ] \right . \Big ] \right. \notag \\
&\left . \left . - \frac{2 k_{0}^{2} r^{2}}{N} \frac{\exp(-2 k_{0} r(N-1)/N) - 1}{\exp(2k_{0} r/N) -1}  - N \exp(-2 k_{0} r) \Phi( \exp(-2k_{0}r/N),2,N) \right . \right . 
\bigg \} \, , \notag
\end{align}
\begin{equation}{\label{eqn:Neta2}}
\langle \eta_{2} \rangle_{D} = c
\end{equation}
\noindent where $H_{(N-1)}^{(2)} = \displaystyle\sum_{n=1}^{N-1} \frac{1}{n^{2}}$ is the generalized harmonic number, $Li_{2}(z) = \displaystyle\sum_{n=1}^{\infty} \frac{z^{n}}{n^{2}}$ is the polylogarithm of index 2 and $\Phi(z,a,b) = \displaystyle\sum_{n=0}^{\infty} \frac{z^{n}}{(b+n)^{a}}$ is the Lerch Phi function.  By taking the large $N$ limit, the above expressions are found to be finite and give  
\begin{equation}{\label{eqn:avgalpha2}}
\langle \alpha_{2} \rangle_{D} = \frac{1}{2} c + \frac{1}{4 k_{0} L^{2}} (1- k_{0} r) - \frac{1}{4 k_{0} L^{2}} (1-k_{0} r) \exp(- 2 k_{0} r) \, ,
\end{equation}
\begin{equation}{\label{eqn:avgdelta2}}
\langle \delta_{2} \rangle_{D} = \frac{1}{2} c - \frac{1}{4 k_{0} L^{2}} (1- k_{0} r) + \frac{1}{4 k_{0} L^{2}} (1-k_{0} r) \exp(-2 k_{0} r) \, ,
\end{equation}
\begin{equation}{\label{eqn:avgeta2}}
\langle \eta_{2} \rangle_{D} = c \, ,
\end{equation}
The disorder averaged solutions $\langle \alpha_{2} \rangle_{D}$ and $\langle \delta_{2} \rangle_{D}$ are divergent in the interior of the geometry as $r \rightarrow \infty$.  Since the gauge field which sources the disorder is a relevant perturbation, this is not surprising.  In order to have a finite solution in the interior, we should resum the perturbative solution in the spirit of the Poincar\'{e}-Lindstedt method for removing secular terms (terms that grow without bound) in perturbative solutions to differential equations.  This kind of situation was also encountered in the case of scalar disorder studied in \cite{Hartnoll2014}.

\section{Resummation of disordered solution}\label{sec:resum}

We have seen that the disorder averaged metric coefficients $\langle \alpha_{2} \rangle_{D}$ (\ref{eqn:avgalpha2}) and $\langle \delta_{2} \rangle_{D}$ (\ref{eqn:avgdelta2}) diverge in the deep interior as $r \rightarrow \infty$ due to the presence of secular terms.   This indicates a break down in the perturbative scheme and needs to be corrected.  This outcome was not totally unexpected, as a similar sort of divergence in one of the disorder averaged metric components was found \cite{Hartnoll2014}.  This divergence was corrected by resumming the disordered solution in the spirit of the Poincar\'{e}-Lindstedt method.  We will adapt this technique to our problem and see that a resummed solution, which has $\langle \alpha_{2} \rangle_{D}$ and $\langle \delta_{2} \rangle_{D}$ finite in the deep interior, is available. 

The procedure is as follows.  We look for additional terms that could contribute at the correct order in the disorder strength, $\overline{V}^{\, 2}$, as they should only correct the unphysical secular terms in the disorder averaged metric.   Furthermore, these new terms must not violate the asymptotically AdS condition imposed on the uncorrected disordered geometry.  The simplest ansatz then for corrected backreacted metric in Fefferman-Graham coordinates is
\begin{equation}{\label{eqn:ResumMetric}}
ds^{2} = \frac{L^{2}}{r^{2}} \left [ - \frac{\alpha(x,r)}{\beta_{1}(r)^{W(\overline{V})}} dt^{2} + dr^{2} + \eta(x,r) dx^{2} + \frac{\delta(x,r)}{\beta_{2}(r)^{P(\overline{V})}} dy^{2} \right ] \, .
\end{equation}
The new functions $\beta_{1}(r)$ and $\beta_{2}(r)$ may be chosen to remove the secular terms in $\langle \alpha_{2} \rangle_{D}$ and $\langle \delta_{2} \rangle_{D}$.  Hence, they should only contribute starting at second order.  Their exponents, then, $W(\overline{V})$ and $P(\overline{V})$ should be expanded in powers of the disorder strength as
\begin{align}
&W(\overline{V}) = W_{2} \overline{V}^{2} + W_{4} \overline{V}^{4} + \cdots \label{eqn:W} \, , \\
&P(\overline{V}) = P_{2} \overline{V}^{2} + P_{4} \overline{V}^{4} + \cdots \label{eqn:P} \,
\end{align}
where, as in the usual Poincar\'{e}-Lindstedt procedure, the $W_{i}$ and $P_{i}$ are constant coefficients that we will adjust in order to remove the secular terms in $\langle \alpha_{2} \rangle_{D}$ (\ref{eqn:avgalpha2}) and $\langle \delta_{2} \rangle_{D}$ (\ref{eqn:avgdelta2}).  In order to ensure that the spacetime geometry remains asymptotically AdS, we must enforce the boundary condition $\beta_{1}(r) \rightarrow 1$ and $\beta_{2}(r) \rightarrow 1$ as $r \rightarrow 0$.  Moreover, in order to ensure that the functions $\beta_{1}(r)$ and $\beta_{2}(r)$ have a chance at removing the problematic terms in the disorder averaged metric components, we should have that $\beta_{1}(r)$ and $\beta_{2}(r)$ diverge as $r \rightarrow \infty$.  This is the only way that a new term can compete with (and ultimately regulate) the already divergent secular terms.  

The placement of the new functions $\beta_{1}(r)$ and $\beta_{2}(r)$ in the metric (\ref{eqn:ResumMetric}) is not arbitrary.  Expanded to second order in the disorder strength, the right hand side of the traced Einstein equations (\ref{eqn:Ein}) is left unchanged.  This is sensible since the source of the disorder, namely the gauge field (\ref{eqn:firstordermax}) is not being modified.  Notice also that $\beta_{1}(r)$ and $\beta_{2}(r)$ are only functions of the radial coordinate $r$ as their sole purpose is to correct the secular terms in the averaged metric components $\langle \alpha_{2} \rangle_{D}$ (\ref{eqn:avgalpha2}) and $\langle \delta_{2} \rangle_{D}$ (\ref{eqn:avgdelta2}).  There is no new function associated with the $x$ direction as $\langle \eta_{2}(x,r) \rangle_{D}$ (\ref{eqn:avgeta2}) is already finite everywhere and does not require a correction.   

The off-diagonal $[r,x]$ component of the traced Einstein equations (\ref{eqn:eom1rx}) is also unchanged.  This is crucial, as this component of the traced Einstein equation allows us to solve for the combination $\alpha_{2}(x,r) + \delta_{2}(x,r)$, and so the $x$ dependence in this combination will also not be modified from the original solution.  The only difference brought about by including $\beta_{1}$ and $\beta_{2}$ is the possibility of modifying the overall partial integration functions of the radial coordinate in $\alpha_{2}(x,r)$, $\delta_{2}(x,r)$ and $\eta_{2}(x,r)$.  It is this modification that will allow us to remove the secular terms from the disorder averaged metric components.  

Finding the solution for the metric coefficients in (\ref{eqn:ResumMetric}) to second order in the disorder strength goes through exactly the same way as for the original case.  We look for an expansion of the form
\begin{align*}
&\alpha(x,r) = 1 + \overline{V}^{2} \alpha_{2}(x,r) + \cdots \\
&\eta(x,r) = 1 + \overline{V}^{2} \eta_{2}(x,r) + \cdots \\
&\delta(x,r) = 1 + \overline{V}^{2} \delta_{2}(x,r) + \cdots \, .
\end{align*}
Plugging this expansion into the traced Einstein equations (\ref{eqn:Ein}) yields a set of partial differential equations
\begin{align}
&-2r \partial_{r}^{2} \alpha_{2} + 2 \partial_{2} \delta_{2} + 2 \partial_{r} \eta_{2} - 2r \partial_{x}^{2} \alpha_{2} + 6 \partial_{r} \alpha_{2} - \frac{6 W_{2}}{\beta_{1}} \partial_{r} \beta_{1} -\frac{2 P_{2}}{\beta_{2}} \partial_{r} \beta_{2} \notag \\
& - \frac{2 r W_{2}}{\beta_{1}^{2}} (\partial_{r} \beta_{1})^{2} + \frac{2 r W_{2}}{\beta_{1}} \partial_{r}^{2} \beta_{1}  = -\frac{r^{3}}{L^{2}} \left [ (\partial_{r} H)^{2} + (\partial_{x} H)^{2} \right ] \label{eqn:Resumtt} \, , 
\end{align}
\begin{align}
&2 r \partial_{r}^{2} \eta_{2} - 2 \partial_{r} \alpha_{2} - 2 \partial_{r} \delta_{2} + 2 r \partial_{r}^{2} \alpha_{2} + 2 r \partial_{r}^{2} \delta_{2} - 2 \partial_{r} \eta_{2} 
+ \frac{2 W_{2}}{\beta_{1}} \partial_{r} \beta_{1} + \frac{2 P_{2}}{\beta_{2}} \partial_{r} \beta_{2} \notag \\
&+ \frac{2 r W_{2}}{\beta_{1}^{2}} (\partial_{r} \beta_{1})^{2}  + \frac{2 r P_{2}}{\beta_{2}^{2}} (\partial_{r} \beta_{2})^{2} - \frac{2 r P_{2}}{\beta_{2}} \partial_{r}^{2} \beta_{2} - \frac{2 r W_{2}}{\beta_{1}} \partial_{r}^{2} \beta_{1} = \frac{r^{3}}{L^{2}} \left [ (\partial_{r} H)^{2} - (\partial_{x} H)^{2} \right ] \label{eqn:Resumrr} \, , 
\end{align}
\begin{align}
\partial_{x} \partial_{r}(\alpha_{2} + \delta_{2}) = \frac{r^{2}}{L^{2}} (\partial_{r} H) (\partial_{x} H) \label{eqn:Resumrx} \, ,
\end{align}
\begin{align}
2 \partial_{r} \delta_{2} - 2 r \partial_{x}^{2} \delta_{2} - 2 r \partial_{r}^{2} \eta_{2} &+ 2 \partial_{r} \alpha_{2} + 6 \partial_{r} \eta_{2} - 2 r \partial_{x}^{2} \alpha_{2} - \frac{2 W_{2}}{\beta_{1}} \partial_{r} \beta_{1} - \frac{2 P_{2}}{\beta_{2}} \partial_{r} \beta_{2}  \notag \\
&= \frac{r^{3}}{L^{2}} \left [ (\partial_{r} H)^{2} - (\partial_{x} H)^{2} \right ] \, , \label{eqn:Resumxx}
\end{align}
\begin{align}
2 \partial_{r} \eta_{2} &- 2 r \partial_{x}^{2} \delta_{2} + 2 \partial_{r} \alpha_{2} - 2 r \partial_{r}^{2} \delta_{2} + 6 \partial_{r} \delta_{2} - \frac{6 P_{2}}{\beta_{2}} \partial_{r} \beta_{2} - \frac{2 W_{2}}{\beta_{1}} \partial_{r} \beta_{1} \notag \\
&  + \frac{2 r P_{2}}{\beta_{2}} \partial_{r}^{2} \beta_{2} - \frac{2 r P_{2}}{\beta_{2}^{2}} (\partial_{r} \beta_{2})^{2}= \frac{r^{3}}{L^{2}} \left [ (\partial_{r} H)^{2} + (\partial_{x} H)^{2} \right ]  \, . \label{eqn:Resumyy}
\end{align}
Notice, the $[r,x]$ component (\ref{eqn:Resumrx}) is unchanged compared to the original $[r,x]$ component (\ref{eqn:eom1rx}), as promised.  In fact, the new traced Einstein equations differ only from the original ones (\ref{eqn:eom1tt}) - (\ref{eqn:eom1yy}) by the addition of terms with derivatives of $\beta_{1}$ and $\beta_{2}$.  In particular, this means that the $x$ dependence of the metric coefficient $\alpha_{2}(x,r)$, $\eta_{2}(x,r)$ and $\delta_{2}(x,r)$ will be unchanged.  Again, this had to be the case since the source for the disorder, namely the gauge field (\ref{eqn:firstordermax}) has not be modified, nor have the right hand sides of the traced Einstein equations (\ref{eqn:Ein}).

The solutions for $\alpha_{2}(x,r)$, $\eta_{2}(x,r)$ and $\delta_{2}(x,r)$ are found to satisfy
\begin{align}
&\alpha_{2}(x,r) + \delta_{2}(x,r) =  G_{1}(x) + G_{2}(r) + \frac{1}{8 L^{2}} \displaystyle\sum_{n=1}^{N-1} \frac{A_{n}^{2}}{k_{n}^{2}} \left ( 1 + 2 k_{n} r + 2 k_{n}^{2} r^{2} \right ) e^{-2k_{n} r} \notag \\
&+ \frac{1}{L^{2}} \displaystyle\sum_{n=1}^{N-1} \displaystyle\sum_{m=1}^{n-1} \frac{A_{n} A_{m} k_{n} k_{m}}{(k_{n} + k_{m})^{4}} \left [ (k_{n} + k_{m})^{2} r^{2} + 2 r (k_{n} + k_{m}) + 2 \right ] \notag \\
& \hskip 15em \times \cos((k_{n} + k_{m})x + \theta_{n} + \theta_{m}) e^{-(k_{n} + k_{m}) r} \label{eqn:ResumQ1} \, ,
\end{align}
where $G_{1}(x)$ is an arbitrary function of $x$ and
\begin{equation}{\label{eqn:ResumG2}}
G_{2}(r) = \tilde{G}_{2}(r) - \frac{1}{16 L^{2}} \displaystyle\sum_{n=1}^{N-1} \frac{A_{n}^{2}}{k_{n}^{2}} \left ( 1 + 2 k_{n} r + 2 k_{n}^{2} r^{2} \right ) e^{-2 k_{n} r} \, .
\end{equation}
The condition (\ref{eqn:ResumQ1}) along with the partial integration function $G_{2}(r)$ (\ref{eqn:ResumG2}) are also true for the original disordered metric solutions up to a new, potentially different, function $\tilde{G}_{2}(r)$.  It satisfies
\begin{align}
r \partial_{r}^{2} \tilde{G}_{2}(r) - 2 \partial_{r} \tilde{G}_{2}(r) &= \frac{r W_{2}}{\beta_{1}} \partial_{r}^{2} \beta_{1} + \frac{r P_{2}}{\beta_{2}} \partial_{r}^{2} \beta_{2} - \frac{2 P_{2}}{\beta_{2}} \partial_{r} \beta_{2} \notag \\
&- \frac{2 W_{2}}{\beta_{1}} \partial_{r} \beta_{1} - \frac{r W_{2}}{\beta_{1}^{2}} (\partial_{r} \beta_{1})^{2} - \frac{r P_{2}}{\beta_{2}^{2}} (\partial_{r} \beta_{2})^{2} \, . \label{eqn:ResumG2Cond} 
\end{align}
This is the new piece that will allow us to tame the secular terms in the disorder averaged solutions.  

The equations of motion (\ref{eqn:Resumtt}) - (\ref{eqn:Resumyy}) may also be combined to get
\begin{align}
&\alpha_{2}(x,r) - \delta_{2}(x,r) = \tilde{G}_{4}(r) - \frac{1}{8 L^{2}} \displaystyle\sum_{n=1}^{N-1} \frac{A_{n}^{2}}{k_{n}^{2}} + \frac{1}{8 L^{2}} \displaystyle\sum_{n=1}^{N-1} \frac{A_{n}^{2}}{k_{n}^{2}} \left ( 1 + 2 k_{n} r + 2 k_{n}^{2} r^{2} \right ) e^{-2 k_{n} r} \notag \\
&+ \frac{1}{4 L^{2}} \displaystyle\sum_{n=1}^{N-1} \displaystyle\sum_{m=1}^{n-1} \frac{A_{n} A_{m}}{k_{n} k_{m}} \left [ 1 + (k_{n} + k_{m}) r + 2 k_{n} k_{m} r^{2} \right ] \cos((k_{n} - k_{m})x + \theta_{n} - \theta_{m}) e^{-(k_{n} + k_{m})r} \notag \\
&-\frac{1}{4 L^{2}} \displaystyle\sum_{n=1}^{N-1} \displaystyle\sum_{m=1}^{n-1} \frac{A_{n} A_{m}}{k_{n} k_{m}} \left [ 1 + (k_{n} + k_{m}) r \right ] \cos((k_{n}-k_{m})x + \theta_{n} - \theta_{m}) e^{-(k_{n} - k_{m}) r} \label{eqn:ResumQ2} \, ,
\end{align}
where the function $\tilde{G}_{4}(r)$ plays an analogous role to $\tilde{G}_{2}(r)$ in (\ref{eqn:ResumQ1}), it satisfies
\begin{align}
r \partial_{r}^{2} \tilde{G}_{4}(r) - 2 \partial_{r} \tilde{G}_{4}(r) &= -\frac{2 W_{2}}{\beta_{1}} \partial_{r} \beta_{1} + \frac{2 P_{2}}{\beta_{2}} \partial_{r} \beta_{2} - \frac{r W_{2}}{\beta_{1}^{2}} (\partial_{r} \beta_{1})^{2} \notag \\
&+ \frac{r P_{2}}{\beta_{2}^{2}}(\partial_{r} \beta_{2})^{2} + \frac{r W_{2}}{\beta_{1}} \partial_{r}^{2} \beta_{1} - \frac{r P_{2}}{\beta_{2}} \partial_{r}^{2} \beta_{2} \, . \label{eqn:ResumG4} 
\end{align}
Finally, the solution for $\eta_{2}(x,r)$ is
\begin{align}
\eta_{2}(x,r) &= \frac{1}{4} r^{2} \partial_{x}^{2} G_{1}(x) + G_{3}(x) - \tilde{G}_{6}(r) - \frac{1}{32 L^{2}} \displaystyle\sum_{n=1}^{N-1} \frac{A_{n}^{2}}{k_{n}^{2}} \left (2 k_{n} r + 1 \right ) e^{-2 k_{n} r} \notag \\
&+ \frac{1}{L^{2}} \displaystyle\sum_{n=1}^{N-1} \displaystyle\sum_{m=1}^{n-1} \frac{A_{n} A_{m}}{(k_{n} + k_{m})^{4}} \left [ (k_{n} + k_{m}) r + 1 \right ] \cos((k_{n} + k_{m}) x + \theta_{n} + \theta_{m}) e^{-(k_{n} + k_{m}) r} \notag \\
& + \frac{1}{16 L^{2}} \displaystyle\sum_{n=1}^{N-1} \frac{A_{n}^{2}}{k_{n}^{2}} \left ( 2 k_{n} r + 1 \right ) \cos^{2}(k_{n} x + \theta_{n}) e^{-2 k_{n} r} \label{eqn:Resumeta2} \, ,
\end{align}
where
\begin{equation}{\label{eqn:ResumG6}}
\tilde{G}_{6}(r) =  \displaystyle\int dr \left [ \frac{1}{r} r \partial_{r}^{2} \tilde{G}_{2} + \frac{r W_{2}}{2 \beta_{1}^{2}} (\partial_{r} \beta_{1})^{2} + \frac{r P_{2}}{2 \beta_{2}^{2}} (\partial_{r} \beta_{2})^{2} - \frac{r P_{2}}{2 \beta_{2}} \partial_{r}^{2} \beta_{2} - \frac{r W_{2}}{2 \beta_{1}} \partial_{r}^{2} \beta_{1} \right ] \, .
\end{equation}
Using these solutions, we will be able to write down a resummed backreacted metric, devoid of secular terms under disorder averaging.

\subsection{Resummed disorder average}\label{sec:resumavg}

With the solutions (\ref{eqn:ResumQ1}), (\ref{eqn:ResumQ2}) and (\ref{eqn:Resumeta2}) at hand, we can impose asymptotically AdS boundary conditions and require that the secular terms in the disorder average vanish.  To do this, we need a choice for $\beta_{1}(r)$ and $\beta_{2}(r)$.  A natural choice is
\begin{equation}{\label{eqn:beta12}}
\beta_{1}(r) = \beta_{2}(r) = \exp \left( \frac{r}{4 L^{2}} \right ) \, ,
\end{equation}
with $W_{2} = 1 = - P_{2}$.  $\overline{V}$ has units $\rm{Length}^{1/2}$, so to second order $\beta_{1}^{W_{2} \overline{V}^{2}}$ and $\beta_{2}^{P_{2} \overline{V}^{2}}$ are dimensionless.  With this choice, the partial integration functions $\tilde{G}_{2}(r)$ (\ref{eqn:ResumG2}) and $\tilde{G}_{4}(r)$ (\ref{eqn:ResumG4}) simplify to
\begin{equation}{\label{G2Simple}}
\tilde{G}_{2}(r) = d_{1} + \frac{1}{3} d_{2} r^{3} \, ,
\end{equation}
and
\begin{equation}{\label{G4Simple}}
\tilde{G}_{4}(r) = d_{3} + \frac{1}{3} d_{4} r^{3} + \frac{1}{2 L^{2}} r \, ,
\end{equation}
where $d_{1}$, $d_{2}$, $d_{3}$ and $d_{4}$ are constants.  

Since $\tilde{G}_{2}(r)$ appears in $\alpha_{2} + \delta_{2}$ and $\tilde{G}_{4}(r)$ appears in $\alpha_{2} - \delta_{2}$, the two metric coefficients will be modified from their original values by
\begin{equation}{\label{eqn:Inalpha2}}
\frac{1}{2} (\tilde{G}_{2} + \tilde{G}_{4}) = \frac{1}{2} (d_{1} + d_{3}) + \frac{1}{6} (d_{2} + d_{4}) r^{3} + \frac{1}{4 L^{2}} r \, ,
\end{equation}
in $\alpha_{2}(x,r)$, while
\begin{equation}{\label{eqn:Indelta2}}
\frac{1}{2}(\tilde{G}_{2} - \tilde{G}_{4}) = \frac{1}{2}(d_{1} - d_{3}) + \frac{1}{6} (d_{2} - d_{4}) r^{3} - \frac{1}{4 L^{2}} r \, ,
\end{equation}
appears in $\delta_{2}(x,r)$.  In $\eta_{2}$, $\tilde{G}_{6}$ (\ref{eqn:ResumG6}) becomes simply
\begin{equation}{\label{eqn:Ineta2}}
\tilde{G}_{6}(r) = \displaystyle\int dr \left [ \frac{1}{2} r \partial_{r}^{2} \tilde{G}_{2}(r) \right] = \frac{1}{3} d_{2} r^{3} \, .
\end{equation}
Setting the constant $d_{2} = 0$ recovers the original solution for $\eta_{2}(x,r)$.  As such, $\langle \eta_{2} \rangle_{D}$ is exactly the same as in (\ref{eqn:avgeta2}).  This is precisely as expected, $\langle \eta_{2} \rangle_{D}$ was already finite everywhere and requires no correction.  

Finally, setting $d_{3} = d_{4} = 0$ the corrections to $\langle \alpha_{2} \rangle_{D}$ and $\langle \delta_{2} \rangle_{D}$ are precisely those needed to remove the secular terms.  We finally have
\begin{equation}{\label{ResummedAlpha2}}
\langle \alpha_{2} \rangle_{D} = \frac{1}{2} c + \frac{1}{4 k_{0} L^{2}} - \frac{1}{4 k_{0} L^{2}} (1 - k_{0} r) \exp(-2 k_{0} r) \, ,
\end{equation}
\begin{equation}{\label{ResummedDelta2}}
\langle \delta_{2} \rangle_{D} = \frac{1}{2} c - \frac{1}{4 k_{0} L^{2}} + \frac{1}{4 k_{0} L^{2}} (1 - k_{0} r) \exp(-2 k_{0} r) \, ,
\end{equation}
and $\langle \eta_{2} \rangle_{D}$ is given in (\ref{eqn:avgeta2}). These results constitute the corrected disorder averaged metric coefficients, all of which are now finite everywhere in the bulk.  

Note also that the choice for $W_{2} = - P_{2}$ ensures that there are no curvature divergences anywhere in the bulk.  Calculating $K \equiv R_{\mu \nu \lambda \sigma} R^{\mu \nu \lambda \sigma}$ through order $\overline{V}^{2}$ in the disorder averaged metric gives $K = \rm{const}$.  

With the resummed solution, we can ask about transport properties of the disordered geometry.  In particular, we are interested in understanding how the conductivity is modified from the initially clean AdS case.  We tackle this question in the next section.  

\section{Conductivity}\label{sec:cond}

In order the calculate the conductivity, we need to turn on a perturbation to the gauge field.  We will be interested in computing the conductivity along the disordered direction $x$.  In pure AdS, the zero-momentum conductivity can be computed by turning on $A_{x}(r,t) = a_{x}(r) e^{-i \omega t}$ and computing the retarded holographic Green's function via the linearized bulk equations of motion.  Taking the $\omega \rightarrow 0$ limit gives the DC conductivity which turns out to be a constant.  This result persists even at finite temperature and is due to bulk electric-magnetic duality \cite{Herzog2007}.  The calculation is simplified in the AdS case, since the gauge field perturbation does not contribute the to linearized energy-momentum tensor and so cannot source any new metric perturbations.  

The situation is not so simple when a non-zero background value for another component of the gauge field is turned on, for example in the RN-AdS geometry.  This means that, generally speaking, the perturbation of the gauge field needed to measure the conductivity will couple to background gauge field component and contribute at linear order in the equations of motion.  As a consequence, new metric perturbations are sourced and can lead to complicated linearized equations.  

In the context of holographic lattices, such as those studied in \cite{Horowitz2012}, \cite{Horowitz2012a} and \cite{Horowitz2013a} the problem is magnified.  Virtually everything that can be sourced is and the perturbation equations turn into a complicated set of partial differential equations which must be solved using numerical techniques.  In \cite{Blake2013}, a holographic lattice, sourced by a periodic scalar field was studied analytically in a perturbative expansion about the lattice strength.  Due to the perturbative nature of the lattice, the linearized equations of motion for the gauge field fluctuation turn out to simplify considerably and only a single metric and scalar perturbation are sourced making the system more amenable to analysis.

For our disordered case, the situation does not simplify quite so drastically, even though the disorder strength is assumed to be weak.  The gauge field perturbation $A_{x}$ will mix with the background $A_{t}$, producing terms which are second order in perturbations (i.e. one power of the disorder strength $\overline{V}$ and $A_{x}$).  These will in turn source metric perturbations of the same order and in principle may source metric perturbations along every direction.  We can, nevertheless, still access information about the DC conductivity of the system by employing a technique first proposed in \cite{Donos2014b}.  The idea is to turn on a perturbation that is linear in time.  An analysis in linear response theory implies that the DC conductivity may be directly calculated by taking advantage of conserved quantities in the bulk.  By using the equations of motion to relate the boundary current to the magnitude of the applied electric field, the DC conductivity follows from Ohm's law. In \cite{Donos2015}, this technique was adopted to study the DC conductivity of inhomogeneous holographic lattices at finite temperature.  In section \ref{sec:Finite}, we will discuss how this technique may be applied to holographic disorder with non-vanishing initial charge density.

We consider a perturbation to the gauge field $A_{x} = a_{x}(r) - Et$, where $E$ is the constant magnitude of the electric field in the $x$ direction.  At the level of the linearized equations of motion, this perturbation may further source metric perturbations $\left \{ \delta g_{tt}, \, \, \delta g_{xx}, \, \delta g_{yy}, \, \delta g_{xt} \right \}$. We have already elected to work in Fefferman-Graham gauge when constructing the backreacted spacetime in order to ensure an asymptotically AdS solution.  As such, we will impose that the metric perturbations obey the usual Fefferman-Graham expansion near the boundary.  In other words, as $r \rightarrow 0$, $\delta g_{a b} = (L^{2}/r^{2}) \delta g_{a b (0)}(x) + \delta g_{a b (1)}(x) + \mathcal{O}(r^{2})$.  Note that the metric perturbations will be at most functions of $x$ and $r$ and will be composed of combinations of periodic functions in $x$.  This is because the metric perturbations are being sourced by terms made up of the gauge field perturbation and the original time component of the gauge field that sources the disorder, so there is no possibility to source any other kind of dependence.  

Using the bulk Maxwell equations $\nabla_{\mu} F^{\mu \nu} = 0$, so in particular then the $x$ and $r$ components require that 
\begin{equation}{\label{eqn:conserveJ}}
\partial_{x} \left ( \sqrt{-g} F^{x r} \right ) = \partial_{r} \left (\sqrt{-g} F^{x r} \right ) = 0 \, .
\end{equation}
Hence, the current $J^{x} \equiv \sqrt{-g} F^{x r}$ is a conserved quantity in the bulk.  As $r \rightarrow 0$, this defines the boundary current in the $x$ direction.  

The conductivity can be read off from Ohm's law, namely $J^{x} = \sigma E$.  Therefore, if we can find an expression for $J^{x}/E$ and take the disorder average of the result, we will have found the DC conductivity directly (up to an overall normalization constant set by the action (\ref{eqn:action})). 

Our strategy will be as follows.  First, we write down an expression for $J^{x}$ linearized about the gauge field and metric perturbations.  Since this quantity is a constant, we can evaluate it anywhere in the bulk. A particularly convenient choice is at the boundary $r = 0$, where, in keeping with our choice of Fefferman-Graham gauge, we enforce asymptotic AdS falloffs for all of the perturbations.  We then take advantage of our perturbative handle, the disorder strength $\overline{V}$.  The linearized equations of motion generally contain terms at many orders in $\overline{V}$.  We start with the leading order Maxwell equations which we solve for $a_{x}(r)$.  This solution is then fed into the next order in perturbations which couples $a_{x}(r)$ to order $\overline{V}$ terms.  These source the metric perturbations, which may be solved for via the Einstein equations and expressed in terms of the magnitude of the applied electric field $E$.  The results solve for $\sigma$ up to second order, after taking the disorder average. The linearized equations of motion contain higher order terms as well, such as $\overline{V}^{2} \delta g_{a b}$.  Since the metric fluctuations are already sourced by terms of the form $\overline{V} a_{x}$, these terms are already fourth order in perturbations and not relevant.   

The linearized expression for $J^{x}$, expanded near the boundary at $r=0$ is
\begin{equation}{\label{eqn:Jxexpand}}
J^{x} = \sqrt{\frac{\delta_{(0)}}{\alpha_{(0)} \eta_{(0)}}} \left [ \alpha_{(0)} a_{x (1)} + \overline{V} H_{(1)} \delta g_{t x (0)} \right ] \, ,
\end{equation}
where the gauge field perturbation has been expanded near the boundary $a_{x} (r) = a_{x (0)} + r a_{x (1)} + \cdots$.  Also 
\begin{align*}
&\alpha(x,r) = \alpha_{(0)}(x) + r \alpha_{(1)}(x) + \cdots \\
&\eta(x,r) = \eta_{(0)}(x) + r \eta_{(1)}(x) + \cdots \\
&\delta(x,r) = \delta_{(0)}(x) + r \delta_{(1)}(x) + \cdots \, ,
\end{align*}
near $r=0$.  The function $H(x,r)$ is related to the disorder source as $A_{t}(x,r) = \overline{V} H(x,r)$ and it has also been expanded near $r=0$ as
\begin{equation}{\label{eqn:Hexpand}}
H(x,r) = H_{(0)}(x) + r H_{(1)}(x) + \cdots \, .
\end{equation}
The next step is to get an expression for the gauge perturbation $a_{x}(r)$.  To do this, we use the linearized Maxwell equations to leading order in perturbations.  This is just the Maxwell equation in AdS, so the solution is easy to find
\begin{equation}{\label{eqn:axSol}}
a_{x}(r) = b_{1} + b_{2} r \, ,
\end{equation}
Where $b_{1}$ and $b_{2}$ are constants.  This solution then couples to the disorder source $A_{t}$ and the original background metric coefficients to produce the metric fluctuations.  In principle, the gauge field fluctuation itself will then receive corrections from the metric perturbations, but this requires going to, at a minimum, third order in perturbation theory.  The upshot is that we can use the baseline solution for $a_{x}$ to extract the DC conductivity to second order in perturbations.  The constant $b_{2}$ can be fixed in terms of the applied electric field $E$.  To see how, replace the time coordinate $t$ with the ingoing coordinate $v = t -r + \mathcal{O}(\overline{V}^{2})$.  Then the full gauge field perturbation $A_{x} = a_{x} - Et$ is a (finite) ingoing solution provided that $b_{2} = E$ up to second order.  This is exactly the condition required in pure AdS to recover the correct DC conductivity and it will ensure that our result contains the pure AdS case plus a possible correction, so that the final disorder averaged solution will take the form
\begin{equation}{\label{eqn:ConductForm}}
\langle \sigma \rangle_{D} = \sigma_{\rm{AdS}} + \overline{V}^{2} \langle \sigma_{\rm{Disorder}} \rangle_{D} \, .
\end{equation}
That is, the original constant AdS DC conductivity plus a correction. 

With $J^{x}$ in (\ref{eqn:Jxexpand}) we must find an expression for $\delta g_{tx(0)}$ in order to get the conductivity.  This can be accomplished via the linearized equations of motion.  Despite being linearized, the equations of motion are still quite complex and finding a solution is a difficult task.  This difficulty may be circumvented by making use of of perturbative expansion in the disorder strength.  To get an expression for $\delta g_{tx(0)}$ it suffices to solve the linearized Einstein and Maxwell equations up to second order in perturbations.  That is, we keep terms of order $a_{x}$, $\overline{V}^{2}$,  $\overline{V}a_{x}$ and $\delta g$. 
 
We have already solved the linearized Maxwell equations through second order in (\ref{eqn:axSol}).  No further corrections to the Maxwell equations are sourced until at least third order in perturbations.  The linearized Einstein equations work out similarly.  There are no contributions to the linearized equations at first order in perturbations.  This had to be the case as all of the metric coefficients and metric perturbations are at least second order perturbative terms.  Every diagonal component of the linearized Einstein equations is just the original equation plus a correction due to a term proportional to $\overline{V} E$ which, in principle could source metric perturbations along these directions.  Remarkably, none of the diagonal components of the linearized Einstein equations, nor the $[x,r]$ component contain the metric perturbation $\delta g_{tx}$ that we need to compute the conductivity.  In fact, the relevant metric perturbation decouples completely from the others and only shows up in $[t,r]$ and $[t,x]$ components.  The equations are
\begin{align}{\label{eqn:gtxeqns}}
&r \partial_{x} \partial_{r} \delta g_{tx} + 2 r \partial_{r} \delta g_{tx} + \overline{V} r (\partial_{x} H) E = 0 \, , \\
&2 \delta g_{tx} - r^{2} \partial_{r}^{2} \delta g_{tx} - 2 r \partial_{r} \delta g_{tx} - \overline{V} r^{2} (\partial_{r} H) E = 0 \, ,
\end{align}
with a solution
\begin{equation}{\label{eqn:gtx}}
\delta g_{tx}(x,r) = \frac{\overline{V} E}{r^{2}} \displaystyle\sum_{n=1}^{N-1} \frac{A_{n}}{k_{n}^{3}} \left ( 2 + 2 k_{n} r + k_{n}^{2} r^{2} \right ) \cos(k_{n} x + \theta_{n}) \exp(-k_{n} r) + 2 \frac{\overline{V} E}{r^{2}} K(x) \, ,
\end{equation}
where $K(x)$ is a function of $x$ that will be fixed momentarily.  As promised, the metric fluctuation is sourced by a term that goes like $\overline{V}$ times a gauge field fluctuation. With (\ref{eqn:gtx}), we can pick off $\delta g_{tx(0)}$ and get and expression for $J^{x}$ in terms of $E$.  We now have all of the ingredients we need to get to the DC conductivity.  Inserting the results back into (\ref{eqn:Jxexpand}), expanding to second order in the disorder strength
\begin{align}
\frac{J^{x}}{E} &= 1 + \frac{\overline{V}^{2}}{2}(\alpha_{2(0)} + \delta_{2(0)} - \eta_{2(0)} )  - \frac{2 \overline{V}^{2}}{L^{2}} \displaystyle\sum_{n=1}^{N-1} \frac{A_{n}^{2}}{k_{n}^{2}} \cos^{2}(k_{n} x + \theta_{n}) \label{eqn:JoverE} \\
& - \frac{2 \overline{V}^{2}}{L^{2}} \displaystyle\sum_{\substack{n,m=1 \\ m \neq m}}^{N-1} \frac{A_{n} A_{m} k_{m}}{k_{n}^{3}} \cos(k_{n} x + \theta_{n}) \cos(k_{m} x + \theta_{m}) + 2 \frac{\overline{V}^{2}}{L^{2}} H_{(1)} K(x) \, , \notag
\end{align}
where, as $r \rightarrow 0$
\begin{align*}
&\alpha(x,r) = 1 + \overline{V}^{2} \alpha_{2}(x,r) \rightarrow 1 + \overline{V}^{2} \left ( \alpha_{2(0)} + r \alpha_{2 (1)} + \cdots \right ) \, , \\
&\eta(x,r) = 1 + \overline{V}^{2} \eta_{2}(x,r) \rightarrow 1 + \overline{V}^{2} \left ( \eta_{2(0)} + r \eta_{2 (1)} + \cdots \right ) \, , \\
&\delta(x,r) = 1 + \overline{V}^{2} \delta_{2}(x,r) \rightarrow 1 + \overline{V}^{2} \left ( \delta_{2(0)} + r \delta_{2 (1)} + \cdots \right ) \, .
\end{align*}
From the metric coefficient $\alpha_{2}$, $\eta_{2}$ and $\delta_{2}$ we get 
\begin{align*}
\alpha_{2 (0)} + \delta_{2 (0)} - \eta_{2 (0)}& = \frac{1}{16 L^{2}} \displaystyle\sum_{n=1}^{N-1} \frac{A_{n}^{2}}{k_{n}^{2}} \cos^{2}(k_{n} x + \theta_{n}) - \frac{1}{32 L^{2}} \displaystyle\sum_{n=1}^{N-1} \frac{A_{n}^{2}}{k_{n}^{2}} \\
& + \frac{1}{L^{2}} \displaystyle\sum_{n=1}^{N-1} \displaystyle\sum_{m=1}^{n-1} \frac{A_{n} A_{m} k_{n} k_{m}}{(k_{n} + k_{m})^{4}} \cos((k_{n} + k_{m})x + \theta_{n} + \theta_{m}) \, .
\end{align*}
The last step needed to get the DC conductivity is to take the disorder average (\ref{eqn:average}) of (\ref{eqn:JoverE}).  To do this, we need to know the function $K(x)$ appearing in (\ref{eqn:gtx}).  This function may be fixed by requiring a finite large $N$ limit for the conductivity as well as a finite value for $\delta g_{tx (0)}$ and a vanishing disorder average for $\delta g_{tx}$.  The condition on the disorder average of $\delta g_{tx}$ actually follows from the requirement of a finite large $N$ limit for the conductivity.  If $\langle \delta g_{tx} \rangle_{D}$ was a constant as $r \rightarrow 0$, then the $H_{(1)} \delta g_{tx(0)}$ term in $J^{x}$ (\ref{eqn:Jxexpand}) would vanish under disorder averaging and would not be able to regulate the large $N$ limit.  The coefficient of $K(x)$ can then be fixed by dimensional analysis.  A good choice is 
\begin{equation}{\label{eqn:Kx}}
K(x) = \frac{\pi^{2}}{12 \zeta(3)}\displaystyle\sum_{n=1}^{N-1} \frac{A_{n}^{3}}{k_{n}^{4}} \cos(k_{n} x + \theta_{n}) \, ,
\end{equation}
where $\zeta(n)$ is the Riemann zeta function.  There is more than one choice for the coefficient of $K(x)$, however every choice which is dimensionally consistent and satisfies our basic requirements for $\delta g_{tx}$ returns the same result for the disorder averaged conductivity.  Applying (\ref{eqn:average}) to (\ref{eqn:JoverE}) and restoring the units (which is due to the overall normalization we have chosen for the action (\ref{eqn:action}) \footnote{In pure AdS, the dual DC conductivity is $\sigma_{DC} = 1/e^{2}$, where the bulk gauge field is normalized by $-1/4e^{2}$.  In our conventions, $e$ has been absorbed into the gauge field and we have pulled out an overall factor of $16 \pi G_{N}$.}) we get
\begin{equation}{\label{eqn:condAvg}}
\langle \sigma \rangle_{D} = \frac{1}{16 \pi G_{N}} + \frac{\overline{V}^{2}}{4 \pi G_{N} L^{2} k_{0}} \, .
\end{equation}
The disorder averaged conductivity takes the form (\ref{eqn:ConductForm}) as claimed.  The overall contribution due to disorder is a constant which comes in a second order in the disorder strength.  This is sensible as the disorder induced corrections to the geometry (\ref{eqn:ResumMetric}) are at second order.  In fact, our result echoes that of the single holographic lattice constructed in \cite{Chesler2014}.  From the point of view of the conductivity, only the metric coefficients $\alpha(x,r)$, $\delta(x,r)$ and $\eta(x,r)$ really matter.  Under disorder averaging, these functions approach a constant towards the boundary $r \rightarrow 0$ which scales as $\overline{V}^{2}/(4 k_{0} L^{2})$, which is a natural dimensionless constant made up of the three scales in the system.  It is therefore not surprising that the correction induced by the disorder source on the DC conductivity involves precisely this combination.  Moreover, notice that as $\overline{V} \rightarrow 0$, the average DC conductivity (\ref{eqn:condAvg}) reduces to the expected pure AdS result.   

Note that the result (\ref{eqn:condAvg}) is specific to a disordered source with a Gaussian distribution (\ref{eqn:Gaussian}).   It is an interesting question as to whether or not changing the disorder distribution will affect the final result.  In particular, the fact that the correction to the conductivity is positive and not negative as might be expected is a result of having chosen a Gaussian distribution. It is unclear that this will be the case if the disorder distribution is modified.  Moreover, it is also unclear that a positive correction persists to higher orders in the disorder strength. It might be that further corrections come in with the opposite sign, but are only visible at larger disorder strength. 

It is tempting to try and address the question of using a different distribution starting from (\ref{eqn:JoverE}). However, to arrive at this formula, we have explicitly made use of the resummed metric (\ref{eqn:ResumMetric}) which implicitly assumes a Gaussian distribution for the disorder source.  In particular, the functions $\beta_{1}(r)$ and $\beta_{2}(r)$ are specific for this realization of disorder.  Understanding how the form of the metric coefficients in the backreacted geometry is dependent on how disorder is implemented in the system and how this affects the dual transport properties are questions we leave for future work.   

\section{Finite charge density}\label{sec:Finite}

The observations of the previous section all follow from a system with zero initial charge density.  Turning on a non-zero background charge density can change the game drastically.  The baseline, clean, geometry is now the Reissner-Nordstr\"{o}m-AdS (RN-AdS) solution.  In the RN-AdS background, the boundary directions preserve translational invariance, and so there is no way to dissipate momentum.  Coupled with the fact that the solution no longer possess particle-hole symmetry like the pure AdS case, the net result is a divergent DC conductivity.  From the perspective of our previous results in pure AdS, it is a natural question to ask how much we can say about adding disorder to this background.  

The action and equations of motion are the same as in (\ref{eqn:action}), (\ref{eqn:Ein}) and (\ref{eqn:Max}).  In this case, the initially clean geometry is Reissner-Nordstr{\"o}m-AdS and the gauge field is not constant everywhere.  The baseline solution is
\begin{equation}{\label{RNAdS}}
ds^{2} = \frac{L^{2}}{r^{2}} \left ( -f(r) dt^{2} + \frac{dr^{2}}{f(r)} + dx_{i}^{2} \right ) \, .
\end{equation}
There are $d-1$ boundary directions.  The boundary is located at $r=0$ in these coordinates. $f(r)$ is given by
\begin{equation}
f(r) = 1 - \left (1 + \frac{(d-2) \mu_{0}^{2} r_{0}^{2}}{2 (d-1) L^{2}} \right ) \left ( \frac{r}{r_{0}} \right )^{d} + \frac{(d-2) \mu_{0}^{2} r_{0}^{2}}{2 (d-1) L^{2}} \left ( \frac{r}{r_{0}} \right )^{2(d-1)} \, ,
\end{equation}
\begin{equation}
A_{t}(r) = \mu_{0} \left ( 1-\left ( \frac{r}{r_{0}} \right )^{d-2} \right ) \, ,
\end{equation}
\noindent where $r_{0}$ is the location of the outer horizon and $\mu_{0}$ is a constant associated to the chemical potential of the dual theory \cite{Hartnoll2009}.  The temperature is
\begin{equation}{\label{temp}}
T = \frac{1}{4 \pi r_{0}} \left ( d - \frac{(d-2)^{2} \mu_{0}^{2} r_{0}^{2}}{2 (d-1) L^{2}} \right ) \, .
\end{equation}
We introduce disorder the same way we did before by adding a perturbation to the gauge field and insisting on the boundary condition (\ref{eqn:chemicalpotential}) and regularity in the interior.   As an example, if $d+1=5$ and we take the limit where all the horizons coincide at $r=r_{0}$ (i.e. $T=0$).  The solution to the Maxwell equations at first order is
\begin{align}
A_{t}(x,r) = \mu_{0} \left ( 1-\frac{r^{2}}{r_{0}^{2}} \right ) + {} &\overline{V} \displaystyle\sum_{n=1}^{N-1} \frac{A_{n}}{3^{\Delta_{n}/2}} \frac{\Gamma(1+\frac{\Delta_{n}}{2}) \Gamma(\frac{\Delta_{n}}{2})}{\Gamma(\Delta_{n})} \left ( 1 + 2 \frac{r^{2}}{r_{0}^{2}} \right ) \left [ \frac{(r_{0}^{2} - r^{2})}{r^{2}} \right ]^{\Delta_{n}/2} \label{eqn:d4T0Sol}  \\
&\times {}_2F_{1} \left [ \frac{\Delta_{n}}{2}, 1 + \frac{\Delta_{n}}{2} ; \Delta_{n} ; \frac{r^{2} - r_{0}^{2}}{3 r^{2}} \right ] \cos(k_{n} x + \theta_{n}) \notag \, ,
\end{align}
where $\Delta_{n} = 1 + \sqrt{1 + \frac{k_{n}^{2} r_{0}^{2}}{3}}$.  Note that the solution (\ref{eqn:d4T0Sol}) vanishes at the horizon $r = r_{0}$ as required for the gauge field to be well defined.  

In $d+1=4$, the solution is more complicated. 
\begin{align}
A_{t}(x,r) = \mu_{0} \left ( 1 - \frac{r}{r_{0}} \right ) + {} &\overline{V} \displaystyle\sum_{n=1}^{N-1} \frac{3^{\lambda_{n}/2} A_{n}}{r_{0}} (r_{0} - r)^{\lambda_{n}/2} \left [ \frac{i \sqrt(2)(r_{0} - r) - 2 (r_{0} + 2 r)}{i \sqrt{2} - 2} \right ]^{1-\lambda_{n}/2} \label{eqn:d3T0Sol} \\
&\times \frac{{}_2F_{1} \left [ \frac{\lambda_{n}}{2}, -\frac{1}{2}; \frac{1+ \lambda_{n}}{2} ; \frac{2 i \sqrt{2}(r-r_{0})}{i \sqrt{2}(r_{0} - r) - 2(r_{0}+2r)} \right ]}{{}_2F_{1} \left [ \frac{\lambda_{n}}{2}, -\frac{1}{2}; \frac{1+ \lambda_{n}}{2} ; -\frac{2 i \sqrt{2}}{i \sqrt{2} - 2} \right ]} \, , \notag
\end{align}
where $\lambda_{n} = 1 + \sqrt{1 + \frac{r_{0}^{2} k_{n}^{2}}{3}}$.  The solution (\ref{eqn:d3T0Sol}) vanishes at the horizon $r =r_{0}$ and obeys the correct boundary condition (\ref{eqn:chemicalpotential}) at $r=0$.  

The idea now is to plug this back into the equations of motion and work out the back reacted geometry to next order in $\overline{V}$ perturbations.  This is more complicated than the initially uncharged case.  Since the initially clean geometry requires a nontrivial gauge field from the get go, there is a baseline term in $A_{t}(x,r)$ that survives on the right hand side of the equations of motion.  In general this will mix perturbative and baseline terms, meaning that the equations of motion must be consistently solved at both first and second order in $\overline{V}$.  Also, the solution presented here is not in the Fefferman-Graham coordinates of the previous section.  This adds the complication that, in general, the backreacted geometry could have off-diagonal $g_{r a}$ contributions.  Since the perturbation is only along the $x$ direction, $A_{t}$ only mixes the $r$ and $x$ directions.  Hence, the only off-diagonal term that can be sourced is $g_{r x}$.  This means that the back-reacted metric should take the form
\begin{equation}{\label{eqn:ChargedMetric}}
ds^{2} = \frac{L^{2}}{r^{2}} \left ( -\alpha(x,r) dt^{2} +\frac{dr^{2}}{F(r)} + \eta(x,r) dx^{2} +  \delta(x,r) dy^{2} + 2 \chi(x,r) dx dr \right ) \, .
\end{equation}
The perturbative expansion for the unknown functions this time is then
\begin{align*}
&\alpha(x,r) = f(z) + \overline{V} \alpha_{1}(x,r) + \overline{V}^{2} \alpha_{2}(x,r) + \cdots \, , \\
&F(r) = f(r) + \overline{V} F_{1}(r) + \overline{V}^{2} F_{2}(r) + \cdots \, , \\ 
&\delta(x,r) = 1 + \overline{V} \delta_{1}(x,r) + \overline{V}^{2} \delta_{2}(x,r) + \cdots \, , \\
&\eta(x,r) = 1 + \overline{V} \eta_{1}(x,r) + \overline{V}^{2} \eta_{2}(x,r)+ \cdots \, , \\
&\chi(x,r) = \overline{V} \chi_{1}(x,r) + \overline{V}^{2} \chi_{2}(x,r) + \cdots \, ,
\end{align*}
\noindent note that the expansion for $\chi(x,r)$ starts at order $\overline{V}$.  This off-diagonal component is not present in the initial clean geometry, so in the limit that $\overline{V} \rightarrow 0$, this metric coefficient must vanish.  Following the same logic as in section \ref{sec:uncharged}, we should try and solve the traced Einstein equations up to second order in the disorder strength.  At zeroth order, the traced Einstein equations are just those of the baseline RN-AdS solution, ensuring that the full solution will be expressed as the RN-AdS geometry plus corrections.  This observation is what constrains the backreacted metric component $g_{rr}$ to be only a function of $r$.  The disorder is turned on along the boundary directions, so we do not expect it to modify the location of the horizon.  At higher order, the traced Einstein equations are complicated and it is unclear if a compact analytic solution can be found.  While we will not attempt to answer this question here, it would be curious to understand if the numerical techniques applied to holographic lattices and scalar disorder could be applied here.  

In lieu of a full analytic solution, we can nevertheless extract the form of the DC conductivity by applying the technique used for inhomogeneous holographic lattices in \cite{Donos2015}.  We will briefly review the salient features of this technique as we go along.  We will make a  modest assumption about the behaviour of the disordered solution near a horizon and see that, given the metric coefficients and gauge field solution, it is possible to extract the form of the disorder averaged DC conductivity directly in terms of horizon data.  

We will assume the form of the backreacted metric is (\ref{eqn:ChargedMetric}) with the event horizon at $r = r_{0}$ and asymptotes to $AdS_{4}$ near the boundary $r = 0$.  As $\overline{V} \rightarrow 0$, the geometry must reduce to the RN-AdS solution.  We will work at finite temperature $T$ and express the gauge field solution as
\begin{equation}{\label{eqn:GaugeFinite}}
A_{t}(x,r) = \mu_{0} \left ( 1-\frac{r}{r_{0}} \right ) + \overline{V} H(x,r) \, ,
\end{equation}
where near the boundary, $H(x,r)$ respects the disordered boundary condition (\ref{eqn:chemicalpotential}).  In general, it will be a sum over periodic functions in $x$ of arbitrary wavelength $k_{n}$. The same must be true of the metric coefficients in (\ref{eqn:ChargedMetric}).  Note that, even though there are many periodicities in the disorder source, and hence the spacetime solution, there is a common periodicity $2 \pi N/k_{0}$.  The metric coefficients and gauge field will also be functions of the random angles $\theta_{n}$ in the disorder source (\ref{eqn:chemicalpotential}). In order for the gauge field to be well defined, it must vanish at the horizon, as in the RN-AdS case.  Hence, the solution (\ref{eqn:GaugeFinite}) should vanish as $r \rightarrow r_{0}$.  

We will insist on an asymptotically AdS solution, so the metric coefficients in the backreacted geometry (\ref{eqn:ChargedMetric}) will need to fall off appropriately near the boundary at $r = 0$.  We will also need to insist on regularity of the solution at the event horizon, $r = r_{0}$.  The metric coefficients in (\ref{eqn:ChargedMetric}) should then be expanded near the horizon as
\begin{align}
&\alpha(x,r) = - 4 \pi T (r - r_{0}) \left [ \alpha_{(0)}(x) + \mathcal{O}((r-r_{0}))  + \cdots \right ] \, , \label{eqn:AlphaNearr0} \\
&F(r) = - 4 \pi T (r - r_{0}) \left [ F_{(0)} + \mathcal{O}((r-r_{0}))  + \cdots \right ]+ \cdots \, , \label{eqn:FNearr0} \\
&\delta(x,r) = \delta_{(0)}(x) + (r- r_{0}) \delta_{(1)}(x) + \mathcal{O}((r-r_{0})^{2})+ \cdots \, , \label{eqn:DeltaNearr0} \\
&\eta(x,r) = \eta_{(0)}(x) + (r- r_{0}) \eta_{(1)}(x) + \mathcal{O}((r-r_{0})^{2})+ \cdots \, , \label{eqn:EtaNearr0} \\
&\chi(x,r) = \chi_{(0)}(x) + (r-r_{0}) \chi_{(1)}(x) + \mathcal{O}((r-r_{0})^{2}) \cdots \, , \label{eqn:ChiNearHr0} \\
&H(x,r) = (r- r_{0}) H_{1}(x) + \mathcal{O}((r-r_{0})^{2}) + \cdots \, , \label{eqn:HNearr0} 
\end{align}  
where each of the functions in the near horizon expansion may themselves be expanded in the disorder strength $\overline{V}$.  For example $\alpha_{(0)} = \alpha_{0(0)} +\overline{V} \alpha_{1(0)}(x) + \overline{V}^{2} \alpha_{2(0)}(x) + \cdots$, and similarly for the rest of (\ref{eqn:AlphaNearr0}) - (\ref{eqn:HNearr0}).  Note that in the limit that $T \rightarrow 0$, there will be terms in $\alpha(x,r)$ and $F(r)$ which are proportional to $T^{-1}$ and stay non-zero.  These terms originate from the background charge density in the initial RN-AdS geometry.  Our convention will be that the first subscript indicates the power of $\overline{V}$ that multiplies the coefficients while the second subscript in brackets indicates the order in the near horizon expansion.  Note also that the function $H(x,r)$ in (\ref{eqn:HNearr0}), which appears in the solution to the gauge field to first order in the disorder strength, only starts at $\mathcal{O}(r-r_{0})$.  This is to ensure that the gauge field vanishes at the horizon, as required by regularity.  To keep the discussion as general as possible, as in \cite{Donos2015}, we turn on every possible perturbation to both the gauge field and the metric, namely $ \left \{ a_{t}(x,r) \, , a_{r}(x,r), \, a_{x}(x,r) \right \}$ and $\left \{ \delta g_{tt} \, , \delta g_{rr} \, , \delta g_{xx} \, , \delta g_{yy} \, , \delta g_{x r} \, , \delta g_{t r} \, , \delta g_{t x} \right \}$.  Note that we do not need to turn on a $y$ component to the gauge field as we will be interested in measuring the conductivity along the disordered $x$ direction.  To this end, we use a linear perturbation
\begin{equation}{\label{eqn:Axtime}}
A_{x}(x,r) = a_{x}(x,r) - Et \, ,
\end{equation}
where $E$ is the constant magnitude of the applied electric field at the boundary.
 
The perturbations will need to fall off appropriately near the boundary so as not to destroy the AdS asymptotics.  Moreover, we will insist on regularity at the horizon $r = r_{0}$.  This condition may be enforced on the perturbations by replacing the time coordinate with the ingoing Eddington-Finkelstein like coordinate
\begin{equation}{\label{eqn:InfallCoord}}
v = t - (4 \pi T)^{-1} \ln(r_{0} - r) + \cdots \, .
\end{equation}
In light of the regularity condition, the constraints on the near horizon behaviour of the gauge field are
\begin{align}
&a_{t}(x,r) = a_{t(0)}(x) + \mathcal{O}(r-r_{0}) + \cdots \label{eqn:atNearr0} \, , \\
&a_{r}(x,r) = \frac{1}{f(r)} \left ( a_{r (0)}(x) +\mathcal{O}(r-r_{0}) + \cdots \right ) \label{eqn:arNearr0} \, , \\
&a_{x}(x,r) = \ln(r_{0} - r) \left ( a_{x(0)} + \mathcal{O}(r-r_{0}) + \cdots \right ) \label{eqn:axNearr0} \, ,  
\end{align}
along with the relations 
\begin{align}
&a_{r(0)} = -a_{t(0)} \label{eqn:arCondition} \, ,  \\
&a_{x(0)} = \frac{E}{4 \pi T} \label{eqn:axCondition} \, ,
\end{align}
each of which follows as a consequence of regularity after switching to the ingoing coordinate (\ref{eqn:InfallCoord}).  Whereas for the metric perturbations, regularity requires
\begin{align}
&\delta g_{tt}(x,r) = \frac{L^{2}}{r_{0}^{2}} f(r) \left ( \delta g_{tt(0)}(x) + \cdots \right ) \label{eqn:gttNearr0} \, , \\
&\delta g_{rr}(x,r) = \frac{L^{2}}{r_{0}^{2} f(r)} \left ( \delta g_{rr(0)}(x) + \cdots \right ) \label{eqn:grrNearr0} \, , \\
&\delta g_{xx}(x,r) = \delta g_{xx(0)}(x) + \mathcal{O}(r-r_{0}) + \cdots \label{eqn:gxxNearr0} \, , \\
&\delta g_{yy}(x,r) =  \delta g_{yy(0)}(x) + \mathcal{O}(r-r_{0}) + \cdots \label{eqn:gyyNearr0} \, , \\
&\delta g_{tr}(x,r) = \delta g_{tr(0)}(x) = \mathcal{O}(r-r_{0}) + \cdots \label{eqn:gtrNearr0} \, , \\
&\delta g_{xr}(x,r) = -\frac{1}{f(r)} \left ( \delta g_{xr(0)}(x) + \mathcal{O}(r-r_{0}) + \cdots \right ) \label{eqn:gxrNearr0} \, , \\
&\delta g_{tx}(x,r) =  \delta g_{tx(0)}(x) + \mathcal{O}(r-r_{0}) + \cdots \label{eqn:gtxNearr0} \, ,
\end{align}
along with the conditions
\begin{align}
&\delta g_{tx(0)} = - \delta g_{xr(0)} \label{eqn:gtxCondition} \, , \\
&\delta g_{tt(0)} + g_{rr(0)} - 2 g_{tr(0)} = 0 \, .
\end{align}
Notice that since the disorder source is a sum over periodic functions in $x$ of arbitrary wavelength, the gauge field and metric perturbations will have to be as well.  Again, they will all share the same common periodicity $2 \pi N/k_{0}$. Furthermore, the perturbations will all be functions of the random angles $\theta_{n}$ in the disorder source (\ref{eqn:chemicalpotential}).   

The next step, as in the case of a single holographic lattice, is to identify two useful conserved quantities in our background.  The first is the boundary current, $J^{x} = \sqrt{-g} F^{x r}$ which, just as in section \ref{sec:cond} is a bulk constant by virtue of the Maxwell equations $\partial_{r} (\sqrt{-g} F^{r x} ) = \partial_{x} (\sqrt{-g} F^{x r}) = 0$.  Again, the gauge field can only be a function of $x$ and $r$. The boundary current, linearized about the gauge field and metric perturbations, is
\begin{equation}{\label{eqn:Chargedbdycurrent}}
J^{x} = \sqrt{\frac{\delta F}{\alpha(\eta - \chi^{2} F)}} \left \{ \alpha \left ( \partial_{x} a_{r} - \partial_{r} a_{x} \right ) + \frac{r^{2}}{L^{2}} \left [ \overline{V} (\partial_{x} H) \delta g_{tr} +\left (\frac{\mu_{0} }{r_{0}} - \overline{V} \partial_{r} H \right ) \delta g_{tx}  \right ] \right \} \, .
\end{equation}
As we are working at finite temperature, there is another conserved quantity, Q, associated the the heat current.  This conserved quantity is associated with a tensor \cite{Donos2014b}
\begin{equation}{\label{eqn:DefineG}}
G^{\mu \nu} = \nabla^{\mu} \xi^{\nu} + \frac{1}{2} \xi^{[\mu}F^{\nu ] \sigma}A_{\sigma} + \frac{1}{4} (\psi - 2 \phi) F^{\mu \nu} \, ,
\end{equation}
where $\xi^{\mu}$ is a Killing vector such that $\mathcal{L}_{\xi} F = 0$.  Also, $\mathcal{L}_{\xi}A = d\psi$ and $i_{\xi} F = d\phi$, $i_{\xi}F$ being the interior product of $\xi$ and $F$.  With these definitions, it can be shown that
\begin{equation}{\label{eqn:derG}}
\nabla_{\mu} G^{\mu \nu} = 3 \xi^{\nu} \, ,
\end{equation}
provided that the Maxwell (\ref{eqn:Max}) and traced Einstein (\ref{eqn:Ein}) equations are satisfied \cite{Donos2014b}. 

An appropriate killing vector is $\xi^{\mu} = [\xi^{t}, \xi^{r}, \xi^{x}, \xi^{y}] = [1,0,0,0]$, which satisfies the requirement $\mathcal{L}_{\xi}F = 0$.  There is a conserved quantity associated with (\ref{eqn:DefineG}) as can be seen by observing that $G^{\mu \nu}$ is antisymmetric and that it is only a function $x$ and $r$.  Hence, $\partial_{x} (\sqrt{-g} G^{r x}) = \partial_{r} (\sqrt{-g} G^{r x}) = 0$, which follows from (\ref{eqn:derG}).  The conserved quantity is $Q = \sqrt{-g}G^{r x}$.  As shown in \cite{Donos2014b}, this quantity is the heat current in the boundary theory.  Linearized about the perturbations, Q is 
\begin{align}
&Q =  \frac{1}{2 r L^{2}} \sqrt{\frac{\delta F}{\alpha ( \eta - \chi^{2} F)}} \left \{ r L^{2} \alpha \left [ \mu_{0} \left (1 - \frac{r}{r_{0}} \right ) + \overline{V} H \right ]  \left ( \partial_{r} a_{x} - \partial_{x} a_{r} \right ) \right . \label{eqn:heatQ} \\
&\left . + r L^{2} \alpha \left ( \partial_{r} \delta g_{tx} - \partial_{x} \delta g_{tr} \right ) + \left ( r^{3} \overline{V} \mu_{0} \left (1 - \frac{r}{r_{0}} \right ) \partial_{r} H  + 2 L^{2} \alpha - \frac{\mu_{0}^{2} r^{3}}{r_{0}} \left (1 - \frac{r}{r_{0}} \right ) - \frac{\mu_{0} r^{3} \overline{V}}{r_{0}} H \right . \right . \notag \\
& \left . \left .  - L^{2} r \partial_{r} \alpha + r^{3} \overline{V}^{2} H \partial_{r} H  \right ) \delta g_{tx} - \left ( L^{2} r \partial_{x} \alpha + \mu_{0} r^{3} \overline{V} \left (1 - \frac{r}{r_{0}} \right )  \partial_{x} H + r^{3} \overline{V}^{2} H \partial_{x} H \right ) \delta g_{tr} \right \} \, . \notag
\end{align}
In what follows, we will make use of the perturbative expansion in the disorder strength and, similarly to the initially uncharged case in section \ref{sec:cond}, we will ignore terms of order $\overline{V}^{2} \delta g$ and higher.  Following \cite{Donos2015}, the next step is to evaluate the constants $J^{x}$ (\ref{eqn:Chargedbdycurrent}) and $Q$ (\ref{eqn:heatQ}) near the event horizon at a fixed temperature $T$.  Using the expansion for the metric coefficients (\ref{eqn:AlphaNearr0})-(\ref{eqn:HNearr0}) and the perturbation in (\ref{eqn:atNearr0})-(\ref{eqn:axNearr0}) and (\ref{eqn:gttNearr0})-(\ref{eqn:gtxNearr0}), we find
\begin{equation}{\label{eqn:JxNearr0}}
J^{x} = \sqrt{\frac{\delta_{(0)} F_{(0)}}{\alpha_{(0)} \eta_{(0)}}} \left [ \alpha_{(0)} \left ( E + \partial_{x} a_{r(0)} \right ) + \frac{r_{0}^{2}}{L^{2}} \left ( \frac{\mu_{0}}{r_{0}} - \overline{V} H \right ) \delta g_{tx(0)} \right ] \, ,
\end{equation}
and
\begin{equation}{\label{eqn:QNearr0}}
Q = -2 \pi T L \delta g_{tx(0)} = \rm{constant} \, ,
\end{equation}
meaning that $\delta g_{tx(0)} = \rm{constant}$.  Next, we expand $Q$ (\ref{eqn:heatQ}) to next order in the near horizon expansion and use the linearized equations of motion to express it entirely in terms of near horizon data.  This expression may then be used to find an expression for $\delta g_{tx(0)}$ in terms of $E$.  The next order expansion of $Q$ is messy and contained in appendix \ref{appendix}.  The result is that
\begin{align}
&\alpha_{(0)} \left ( \mu_{0} - r_{0} \overline{V} H_{(1)} \right ) \left ( E + \partial_{x} a_{r(0)} \right ) - 4 \pi T r_{0} \partial_{x} \left ( \alpha_{(0)} \delta g_{tr(0)} \right )  \label{eqn:QNextorder} \\
&+ \left ( 8 \pi T \alpha_{(0)} + 2 \frac{r_{0}^{2}}{L^{2}} \overline{V} \mu_{0} H_{(1)} - \frac{r_{0} \mu_{0}^{2}}{L^{2}}  + \frac{8 \pi^{2} T^{2} r_{0} \alpha_{(0)} \chi_{(0)}^{2}}{\eta_{(0)}}  - \Omega_{(0)}(x) \right ) \delta g_{tx(0)} = 0 \notag \, ,
\end{align}
where $\Omega_{(0)}(x)$ is a function of metric coefficients near the horizon.  It is found in appendix \ref{appendix}.  To get an expression for $J^{x}$ in terms of $E$ and $\delta g_{tx(0)}$, we integrate (\ref{eqn:QNearr0}) over the common periodicity $2 \pi N/k_{0}$.  Then, doing exactly the same thing with (\ref{eqn:QNextorder}), we can relate $E$ to $\delta g_{tx(0)}$.  Substituting this into the expression for $J^{x}$ gives us the desired relation between $J^{x}$ and $E$.  The final result is
\begin{equation}{\label{eqn:JxoverECharged}}
\frac{J^{x}}{E} = \frac{I_{1}}{I_{2} + I_{3}} \, ,
\end{equation}
where
\begin{equation}{\label{eqn:I1}}
I_{1} = \frac{k_{0}}{2 \pi N} \displaystyle\int_{0}^{2 \pi N/k_{0}} \left [  \frac{2 r_{0} \mu_{0}^{2}}{L^{2}} -  \frac{4 r_{0}^{2} \mu_{0}}{L^{2}} \overline{V} H_{(1)} - 8 \pi T \alpha_{(0)} - \frac{8 \pi^{2} T^{2} r_{0}}{\eta_{(0)}} \alpha_{(0)} \chi_{(0)}^{2} + \Omega_{(0)}(x) \right ] dx \, ,
\end{equation}
\begin{align}
I_{2} &= \frac{k_{0}^{2}}{4 \pi^{2} N^{2}} \displaystyle\int_{0}^{2 \pi N/ k_{0}} \sqrt{\frac{\eta_{(0)}}{\alpha_{(0)} \delta_{(0)} F_{(0)}}} dx \label{eqn:I2} \\ &\times\displaystyle\int_{0}^{2 \pi N/ k_{0}} \left [  \frac{2 r_{0} \mu_{0}^{2}}{L^{2}} -  \frac{4 r_{0}^{2} \mu_{0}}{L^{2}} \overline{V} H_{(1)} - 8 \pi T \alpha_{(0)} - \frac{8 \pi^{2} T^{2} r_{0}}{\eta_{(0)}} \alpha_{(0)} \chi_{(0)}^{2} + \Omega_{(0)}(x) \right ] dx \, , \notag
\end{align}
\begin{align}
I_{3} &=  \frac{k_{0}^{2} r_{0}}{4 \pi^{2} L^{2} N^{2}} \displaystyle\int_{0}^{2 \pi N/ k_{0}} \left [ \alpha_{(0)}^{-1} \left ( \overline{V} r_{0} H_{(1)} - \mu_{0} \right ) \right ] dx \label{eqn:I3} \\
& \times \displaystyle\int_{0}^{2 \pi N/ k_{0}} \sqrt{\frac{\alpha_{(0)} \eta_{(0)}}{\delta_{(0)} F_{(0)}}} \left ( \mu_{0} - r_{0} \overline{V} H_{(1)} \right ) dx \, . \notag
\end{align}
All of the functions in the integrals (\ref{eqn:I1})-(\ref{eqn:I3}) may be expanded to second order in the disorder strength $\overline{V}$, provided the solutions are known.  The final step after the expansion is to take the disorder average (\ref{eqn:average}) of (\ref{eqn:JxoverECharged}).  The disorder averaged conductivity is then $16 \pi G_{N} \langle \sigma \rangle_{D} = \langle J^{x}/E \rangle_{D}$.  
\section{Summary and outlook}\label{sec:discussion}

In section \ref{sec:uncharged} we construct a bottom-up holographic model with perturbatively charged disorder. Starting from a bulk Einstein-Maxwell action, we include disorder in the dual theory by using a spectral representation.  This technique is known to simulate a stochastic process \cite{Shinozuka1991}. This is achieved by including a randomly varying chemical potential (\ref{eqn:chemicalpotential}) made up of a sum of $N$ periodic functions along one of the boundary directions; an approach reminiscent of the disordered holographic superconductors studied in \cite{Arean2014}, \cite{Zeng2013} and \cite{Arean2014a}.  The random chemical potential contains two parameters, a wavenumber  $k_{0}$ which is held fixed in the large $N$ limit and $\overline{V}$ which controls the strength of the disorder.  The parameter $\overline{V}$ is taken to be small and is used as a perturbative handle to construct a bulk solution.  A bulk gauge field is turned on which approaches the fluctuating chemical potential (\ref{eqn:chemicalpotential}) near the spacetime boundary.  By letting the gauge field backreact on the initially clean (i.e. zero disorder) $AdS_{4}$ geometry, we construct an asymptotically AdS solution to the bulk equations of motion (\ref{eqn:Ein}) and (\ref{eqn:Max}) at second order in the disorder strength in section \ref{sec:sol2}.  

We evaluate the disorder average (\ref{eqn:average}) of the second order metric coefficients (\ref{eqn:Nalpha2}), (\ref{eqn:Ndelta2}) and (\ref{eqn:Neta2}), in section \ref{sec:avg} and find that they are compactly expressed in terms of special functions.  By carefully evaluating the large $N$ limit, we find a divergence in the deep interior as $r \rightarrow \infty$ in two of the metric coefficients (\ref{eqn:avgalpha2}) and (\ref{eqn:avgdelta2}), indicating that the solutions must be regulated.

In section \ref{sec:resum}, we resum the disordered solution found in section \ref{sec:uncharged} by adapting the standard Poincar\'{e}-Lindstedt method for regulating perturbative solutions to differential equations, similarly to the case of scalar sourced disorder in \cite{Hartnoll2014}.  We find a regulated, second order, solution to the equations of motion with metric functions (\ref{ResummedAlpha2}), (\ref{ResummedDelta2}) and (\ref{eqn:avgeta2}), where the previous noted divergences are removed.  The averaged resummed solution is devoid of curvature singularities through second order in the disorder strength.

With the resummed solution at hand, we study the resulting DC conductivity along the disordered direction in section \ref{sec:cond}.  We directly access the DC conductivity of the model by adapting a technique first proposed in \cite{Donos2014b}.  The idea is to turn on a source linear in time and take advantage of the existence of conserved quantities in the bulk to find a relationship between the boundary current and the magnitude of the applied electric field. The DC conductivity may then be extracted from Ohm's law.  By virtue of the nature of the disordered spacetime solution in this case, turning on a bulk gauge field perturbation along the disordered direction results in a complicated set of equations of motion.  As is the case for single holographic lattices \cite{Horowitz2012}, the gauge field perturbation further sources a whole set of possible metric fluctuations.  By taking advantage of our perturbative handle, namely the disorder strength, the situation simplifies and the relevant metric fluctuation can be solved via the linearized equations of motion up to second order in perturbations.  The disorder averaged DC conductivity is computed in (\ref{eqn:condAvg}) which is found to be the usual AdS conductivity plus a correction at second order in the disorder strength, a result reminiscent of the single holographic lattices studied in \cite{Chesler2014}.

Section \ref{sec:Finite} makes some observations about adding a disordered chemical potential to an initially clean system with a finite charge density.  In this case, the baseline geometry is the Reissner-Nordstr\"{o}m-AdS (RN-AdS) solution.  The presence of an initial charge density changes the character of the backreacted equations of motion, as the perturbative contributions originating from the disorder source now mix with the baseline gauge field.  The result is that geometry may receive corrections at all orders in the disorder source, unlike the initially uncharged case studied in section \ref{sec:uncharged}.  At second order, the traced Einstein equations are complicated and it is not clear that there is a compact analytic solution.  It is still possible to extract some information about the form of the disorder averaged DC conductivity in this case by applying the techniques of \cite{Donos2014b} and \cite{Donos2015}.  The procedure is similar to the initially uncharged case in section \ref{sec:cond}, except that now the baseline solution (and hence the disordered solution) has an event horizon.  By writing down a  broad ansatz for a $3+1$ dimensional disordered geometry at finite charge density and temperature (\ref{eqn:ChargedMetric}), we show that the disorder averaged DC conductivity may be expressed entirely in terms of near horizon data (\ref{eqn:JxoverECharged}), as is the case for the holographic lattices studied in \cite{Donos2015}.  We leave the difficult task of finding a disordered, finite charge density, spacetime solution for future work.  

There many open questions with regard to explicit implementations of holographic disorder, both in terms of studying the properties of the dual field theory as well as understanding the kinds of bulk geometries that arise in the process.

A natural extension to our work here is to disordered holographic superconductors, such as those studied in \cite{Arean2014}, \cite{Zeng2013} and \cite{Arean2014a}.  A spectral representation for the disordered chemical potential is also used in these studies and the properties of the superconducting transition are studied numerically.  It would be interesting to understand how the inclusion of backreaction of the disorder source onto the spacetime geometry changes the picture here.  For example, how is the appearance of the superconducting phase transition affected? Is the critical temperature significantly changed?  Is possible to get an analytical handle on a a critical amount of holographic disorder beyond which the conductivity becomes completely suppressed and the superconducting phase transition does not occur?  This previous question is particularly pertinent with regards to many-body localization.  If such a transition does occur, what kind of bulk geometry is required and how does it fit into conventional gravitational models?  In particular, if a localization transition does occur in the dual theory, would the bulk probe effectively become stuck\footnote{We thank Omid Saremi for pointing this out to us.} at some radial position?  Would this translate to a well defined mobility edge at the corresponding energy scale in the dual theory? 
It would also be interesting to understand the behaviour of time dependent probes in such a background.   

To fully study this problem, it may be necessary to move beyond the perturbative disorder studied in this paper and an analytical approach may be ill-suited and numerical solutions may be required.  In such a case, it would be interesting to understand if the techniques used in \cite{Hartnoll2014} for scalar disorder and \cite{Hartnoll2014a} for holographic lattices would be useful.  

In \cite{Hartnoll2014}, disorder is sourced by a scalar field in $2+1$ dimensions and the disorder averaged metric is found to display an emergent Lifshitz scaling.  It would interesting to classify the possible IR geometries that can be produced in this way, hopefully leading to a better understanding of disorder fixed points of condensed matter systems.  For example, could an interior geometry with an emergent hyperscaling violation exponent be generated via a back-reacted disordered source?  How about IR geometries that break rotations, i.e. in relation to Bianchi models \cite{Iizuka2012}, \cite{Iizuka2013}? In order words, what kind of IR disordered fixed points can be constructed via holography?

In \cite{Saremi2014}, an ansatz for a disordered geometry is proposed and the dynamics of a scalar field in this background are studied.  Using techniques from random matrix theory, a transition is observed which is reminiscent of a disorder driven metal-insulator transition.  Understanding more than one explicit example of a geometry which displays this behaviour as well as the matter content required to support such solutions in a gravitational theory may shed light on the minimal ingredients necessary for accessing disordered phenomena via holography.  It would also be worthwhile understanding how the proposed spacetime in \cite{Saremi2014} fits in with backreacted disordered geometries.  

Finally, our results for the initially uncharged case in sections \ref{sec:uncharged}, \ref{sec:resum} and \ref{sec:cond} depend sensitively on the disorder distribution.  In this paper we have focused on the effect of Gaussian random disorder.  It would be interesting to extend our results to other distributions and understand how the resulting backreacted geometry and transport properties are modified.

\bigskip
\section*{Acknowledgements}

We wish to thank Omid Saremi for helpful discussions in the beginning stages of this project. 

DKO wishes to thank the Institute for Advanced Study for hospitality during the June 2014 Prospects in Theoretical Physics summer school.  

This research was supported by the Natural Sciences and Engineering Research Council (NSERC) of Canada. The research of DKO was supported in part by an E.C.~Stevens Fellowship.

\newpage

\appendix

\section{Appendix} \label{appendix}

As mentioned in section \ref{sec:Finite}, one of the ingredients needed to compute the DC conductivity is the expansion of the heat current Q to order $r-r_{0}$ near the horizon $r_{0}$.  This results in the following condition 
\begin{align}
&\alpha_{(0)} \left ( \mu_{0} - r_{0} \overline{V} H_{(1)} \right ) \left ( E + \partial_{x} a_{r(0)} \right ) - 4 \pi T r_{0} \partial_{x} \left ( \alpha_{(0)} \delta g_{tr(0)} \right ) \label{QNextOrderFull} \\
&+ \left [ 8 \pi T \alpha_{(0)} + \frac{2 r_{0}^{2} \mu_{0} \overline{V}}{L^{2}} H_{(1)} - \frac{\mu_{0}^{2} r_{0}}{L^{2}} + \frac{8 \pi^{2} T^{2} r_{0}}{\eta_{(0)}} \alpha_{(0)} \chi_{(0)}^{2} \right ] \delta g_{tx(0)} \notag \\
&- \left [ 24 r_{0} \pi^{2} T^{2} \alpha_{(1)}  + 8 \pi^{2} T^{2} r_{0} \alpha_{(0)} \frac{F_{(1)}}{F_{(0)}} + 2 \pi T r_{0} \alpha_{(0)} \frac{\delta_{(1)}}{\delta_{(0)}} \- 2 \pi T r_{0} \alpha_{(0)} \frac{\eta_{(1)}}{\eta_{(0)}} \right ] \delta g_{tx(0)} = 0 \, . \notag
\end{align}
Using the linearized equations of motion, it is possible to solve for last term in brackets containing $\alpha_{(1)}$, $F_{(1)}$, $\delta_{(1)}$, $\eta_{(1)}$ and express it entirely in terms of horizon data with the result (\ref{eqn:QNextorder}) mentioned in the main text.  To the relevant order in perturbations, we get 
\begin{align}
\Omega_{(0)}(x) = {} & \frac{1}{2 L^{2} F_{(0)} \eta_{(0)}^{2} \delta_{(0)}^{2} ( 16 \pi^{2} T^{2} F_{(0)} \alpha_{(0)} \delta_{(0)} - 1)} \left .  \Big\{  2 F_{(0)} \eta_{(0)} \delta_{(0)}^{2} \left [ 128 \pi^{3} T^{3} L^{2} F_{(0)} \alpha_{(0)}^{2} \eta_{(0)} \delta_{(0)} \right . \right . \notag \\
& \left . \left . {} - 8 \pi T L^{2} \alpha_{(0)} \eta_{(0)} - 2 \mu_{0} \overline{V} r_{0}^{2} \eta_{(0)} H_{(1)} + 32 \pi^{2} T^{2} \mu_{0} r_{0}^{2} \overline{V} \alpha_{(0)} F_{(0)} \eta_{(0)} \delta_{(0)} H_{(1)} \right . \right . \notag \\
& \left . \left . {} - 16 \pi^{2} T6{2} \mu_{0}^{2} r_{0} \alpha_{(0)} F_{(0)} \eta_{(0)} \delta_{(0)} + 128 \pi^{4} T^{4} r_{0} F_{(0)}^{2} \alpha_{(0)}^{2} \delta_{(0)} \chi_{(0)}^{2} - 8 \pi^{2} T^{2} L^{2} r_{0} F_{(0)} \alpha_{(0)} \chi_{(0)}^{2} \right ] \right . \notag \\
&\left . {} + 4 \pi T r_{0} L^{2} F_{(0)} \delta_{(0)} \left [ 16 \pi^{2} T^{2} F_{(0)} \alpha_{(0)}^{2} \delta_{(0)}^{2} \chi_{(0)} \partial_{x} \eta_{(0)} + 16 \pi^{2} T^{2} F_{(0)} \alpha_{(0)} \eta_{(0)} \delta_{(0)} \chi_{(0)} \partial_{x} \alpha_{(0)} \right . \right . \notag \\
& \left . \left . {} - 32 \pi^{2} T^{2} F_{(0)} \alpha_{(0)}^{2} \eta_{(0)} \delta_{(0)}^{2} \partial_{x} \chi_{(0)} + 16 \pi^{2} T^{2} \alpha_{(0)}^{2} F_{(0)} \eta_{(0)} \delta_{(0)} \partial_{x} \delta_{(0)} + 2 \alpha_{(0)} \eta_{(0)} \delta_{(0)} \partial_{x} \chi_{(0)} \right . \right . \notag \\
& \left . \left . {} - \alpha_{(0)} \delta_{(0)} \chi_{(0)} \partial_{x} \eta_{(0)} - 3 \eta_{(0)} \delta_{(0)} \chi_{(0)}\partial_{x} \alpha_{(0)} - \alpha_{(0)} \eta_{(0)} \chi_{(0)} \partial_{x} \delta_{(0)} \right ] \right . \notag \\
&\left . {} + 16 \pi^{2} T^{2} L^{2} r_{0} \left [ F_{(0)} \eta_{(0)} \delta_{(0)}^{3} ( \partial_{x} \alpha_{(0)})^{2} + F_{(0)} \alpha_{(0)}^{2} \eta_{(0)} \delta_{(0)} (\partial_{x} \delta_{(0)})^{2} + \alpha_{(0)}^{2} \delta_{(0)}^{2} (\partial_{x} \eta_{(0)}) (\partial_{x} \delta_{(0)}) \right . \right . \notag \\
& \left . \left . {} + F_{(0)} \alpha_{(0)} \delta_{(0)}^{3} (\partial_{x} \alpha_{(0)})(\partial_{x} \eta_{(0)}) \right ] - L^{2} r_{0} \left [ \eta_{(0)} \delta_{(0)} (\partial_{x} \alpha_{(0)})(\partial_{x} \delta_{(0)}) + \alpha_{(0)} \delta_{(0)} (\partial_{x} \eta_{(0)}) (\partial_{x} \delta_{(0)}) \right . \right . \notag \\
& \left . \left .  {} + \alpha_{(0)} \eta_{(0)} (\partial_{x} \delta_{(0)})^{2} \right ] -2 L^{2} r_{0} \alpha_{(0)} \eta_{(0)} \delta_{(0)} \left [ 16 \pi^{2} T^{2} F_{(0)} \delta_{(0)}^{2} \partial_{x}^{2} \alpha_{(0)} + 16 \pi^{2} T^{2} F_{(0)} \alpha_{(0)} \delta_{(0)} \partial_{x}^{2} \delta_{(0)} \right . \right . \notag \\
&\left . \left . {} - \partial_{x}^{2} \delta_{(0)} \right ] \right .\Big\} \notag
\end{align}
%

\newpage
\bibliographystyle{utphys}
\bibliography{Disorderbib}

\end{document}